\documentclass[aps,prc,preprint,showpacs]{revtex4-1}
\usepackage{amsmath}
\usepackage{slashed}
\usepackage{color}
\usepackage{appendix}

\usepackage{multirow}

\usepackage{graphicx}

\begin{document}

\title{A study of the $KN$-$K^*N$ coupled system in s-wave}

\author{K.~P.~Khemchandani$^{1,2}$}
\author{ A.~Mart\'inez~Torres$^1$}
\author{F.~S.~Navarra$^1$}
\author{M.~Nielsen$^1$}
\author{L.~Tolos$^{3,4}$}
 \affiliation{$^1$Instituto de F\'isica, Universidade de S\~ao Paulo, C.P 66318, 05314-970 S\~ao Paulo, SP, Brazil.\\
 $^2$Faculdade de Tecnologia, Universidade do Estado do Rio de Janeiro,
Rod. Presidente Dutra Km 298, P\'olo Industrial, 27537-000 , Resende, RJ, Brazil.\\
$^3$Instituto de Ciencias del Espacio  (IEEC/CSIC), Campus Universitat Aut\`onoma de Barcelona, Facultat de Ci\`encies, Torre C5, Bellaterra, E-08193, Spain.\\
$^4$Frankfurt Institute for Advanced Studies (FIAS), 60438 Frankfurt am Main, Germany}

\begin{abstract}
We study the strangeness $+1$ meson-baryon systems  to obtain improved $KN$ and $K^*N$ amplitudes and  to look for a possible resonance formation by the $KN$-$K^*N$ coupled interaction. We obtain amplitudes for light vector meson-baryon systems by implementing the $s$-, $t$-, $u$- channel diagrams  and a contact interaction. The pseudoscalar meson-baryon interactions are obtained by relying on the Weinberg-Tomozawa theorem. The transition amplitudes between the systems consisting of pseudoscalars and vector mesons are calculated  by extending the Kroll-Ruderman term for pion photoproduction replacing the photon by a vector meson.   We fix the subtraction constants required to calculate the loops by fitting our $KN$ amplitudes to the data available for the isospin 0 and 1  $s$-wave phase shifts.  We provide the scattering lengths and the total cross sections for the $KN$ and $K^* N$ systems obtained in our model, which can be useful in future in-medium calculations. Our amplitudes do not correspond to formation of any resonance in none of the isospin and spin configurations.
\end{abstract}

\maketitle

\section{Introduction}
Recent experimental studies related to the production of the $K$ and $K^*$-mesons in proton-proton and proton-nucleus collisions are bringing forward some intriguing findings which seem to call for a reinvestigation of the interaction of these mesons with  nuclear matter.  For example, a deep sub-threshold $K^{*0}$ production has been reported in Ar+KCl collisions by HADES, with the experimental $K^{*0}$ yield and $K^{*0}/K^0$ being overestimated by about a factor five and two, respectively, when applying the UrQMD transport approach \cite{hades}. Recent results on $K^0$ production reported in $p+p$ collisions at 3.5 GeV by the HADES Collaboration show a dominant resonant production coming from $\Delta(1232)$ and $\Sigma(1385)$ for intermediate energies in the formation of $K^0$~\cite{Agakishiev:2014nim}. The need for a reliable information on the in-medium kaon potential has also been discussed in Ref.~\cite{Agakishiev:2014moo} where the $K^0$ production in $p+$Nb reactions at a beam kinetic energy of 3.5 GeV is analyzed by HADES. Further, by analyzing the freeze-out temperature using a statistical hadronization model, the experimental results for $K^{*0}$ production seem to indicate the necessity of considering the rescattering of the decay products of $K^{*0}$ in the hadronic matter \cite{hades}. The importance of the hadronic interactions of the final states is also realized for the $K^{*0}$ production in Au+Au and Cu+Cu at $\sqrt{s_{NN}}=$62.4 and 200 GeV collisions by the STAR Collaboration \cite{star}. There are some unsolved issues present in the field from some previous experiments too. For example, the attenuation of the $K^*$ and $\bar K^*$ states in the hadronic phase of the expanding fireball, as determined by the observation of a strong suppression of the total yield ratios $<K^*>/<K^+>$ and $<\bar{K}^*>/<K^->$ in central Pb$+$Pb collisions compared to p$+$p, C$+$C and Si$+$Si by the NA49 Collaboration \cite{NA49}, was not reproduced  using UrQMD \cite{modelhic1,modelhic2} or statistical HQGM \cite{Becattini:2005xt} models. These findings  indicate the importance of obtaining a reliable determination of $K$ and $K^*$ interactions with nuclear matter. For such studies, it is  helpful to first have information on (free) $KN$ and $K^*N$ interaction. The motivation of the present paper is to make an attempt to revise the information available on the $KN$ and $K^*N$ s-wave interactions from previous studies by using the more complete approach of Refs.~\cite{vbvb,pbvb,hyperons,more} and by constraining the resulting amplitudes to fit the relevant available data.

In Refs.~\cite{vbvb,pbvb,hyperons,more} a detailed investigation of coupled light meson-baryon systems involving pseudoscalars and vector mesons was performed.  The basic vector meson-baryon (VB) Lagrangian, used  in Refs.~\cite{vbvb,pbvb,hyperons,more}, is based on the hidden local symmetry \cite{bando} which treats vector mesons consistently with the chiral symmetry. A study of VB systems done in Ref.~\cite{vbvb} showed that the gauge invariance of the Lagrangian enforces the consideration of a contact term together with $s$-, $t$-  and $u$-channel interactions, which all turn out to give important contributions. This result shows that the low-energy theorems related to the pseudoscalar mesons cannot be extended to the VB systems, implying  that it is important to solve the Bethe-Salpeter equation using the sum of all these interactions as the kernel.  The formalism was further extended to couple VB with pseudoscalar meson-baryon (PB) systems by extending the Kroll-Ruderman theorem for the  pion photoproduction by replacing the photon by a vector meson  in accordance with  vector meson dominance. This formalism has been used to solve coupled-channel equations for strange and nonstrange meson-baryon systems \cite{hyperons,more}, and has been found to be useful in reproducing some relevant experimental data and understanding the properties of several resonances.

A question might arise at this point: why should coupling pseudoscalar and vector baryons be useful in the light baryon sector?  We would like to recall that efforts in this direction have been made earlier within different formalisms \cite{GarciaRecio:2005hy,pbvb,hyperons,more,juelich,lutz,carmen2,javi,javi2} showing that this coupling is important for reproducing experimental data. For example, it is crucial for calculating some properties of a resonance such as its partial decay widths for different  channels. In some cases this coupling can be important to even obtain the right mass and quantum numbers of a resonance, especially when the thresholds of the channels with pseudoscalars and vectors are closely spaced, like $K\Lambda, K\Sigma$, $\rho N, \omega N$ in the nonstrange sector, $K\Xi$ and $\rho \Lambda$ in the strangeness $-1$ sector. Although the $KN$ and $K^*N$ thresholds are not as close as the preceding examples, as we shall explain in the next section, it is important to couple them since the $KN$ interaction in the isospin 0, spin 1/2 configuration is null.  The coupling to the $K^*N$ system, then,  is useful to obtain non-zero phase shifts and scattering lengths and compare them with those available from the partial wave analyses of the relevant experimental data \cite{prc75,gwu,prc29,martin}. 

It is  also our intention to look for a possible resonance in the $KN$-$K^*N$ coupled channel system. Although the $KN$ interaction is known to be repulsive, it is possible that its coupling to $K^*N$  can result in the formation of a resonance. In fact such a possibility has been explored earlier in a study of $KN$-$K^*N$ coupled systems~\cite{GarciaRecio:2005hy} within a formalism based on a SU(6) spin-flavor symmetry for the light hadrons. A calculation of the Weinberg-Tomozawa term ($t$-channel) for all channels was done in Ref.~\cite{GarciaRecio:2005hy} and as a result an  isoscalar resonance with spin-parity $3/2^-$ and mass between 1.7-1.8 GeV was obtained.  A similar situation has been found  in the case of anticharm meson-baryon systems in Ref.~\cite{Gamermann:2010zz}, where a resonance with spin-parity $1/2^-$ and isospin 0 is obtained when pseudoscalar meson-baryon and vector meson-baryon channels are coupled. In this case the uncoupled amplitudes  are null and the resonance is obtained only as a consequence of coupling the two channels,  i.e., due to the transition amplitude. An anticharm baryon (like a strange baryon) necessarily requires a five quark content \cite{Gamermann:2010zz,Aktas:2004qf,Kim:2004pu,Lee:2005pn,Albuquerque:2013hua}.

It might sound discouraging to look for a strangeness $+1$ resonance with the failure of several experiments in finding  $\Theta^+(1540)$ \cite{leps},  (for a review on this, see Ref.~\cite{rev} and for an alternative explanation for the enhancement of the cross section seen in Ref.~\cite{leps}  see Refs.~\cite{amt1,amt2}). These findings, however, do not imply that no strangeness $+1$ baryon exist. Maybe one has to look for a state with a different mass and width, for instance, closer to the $K^*N$ threshold as indicated in Refs.~\cite{GarciaRecio:2005hy,kelkar,oldpaprs}.  

In fact, the consideration of the possibility of formation of a resonance in the present case is  very much in line with the studies of five-quark or meson-baryon systems with strangeness $-1$ or $0$ \cite{Yuan:2012wz,Zou:2006uh,osetramos,jido,Meissner:1999vr,Doring:2009uc}. In these latter works it has been shown that 
a five valence quark content or a meson-baryon content of the resonances formed in these systems is essentially needed to reproduce the relevant experimental data. In this sense, such strangeness $-1$ or $0$ resonances can also be considered as pentaquark states. Actually similar investigations  have even been extended  to systems of  two-mesons and a baryon where resonances arising purely from the three-body dynamics have been deduced \cite{MartinezTorres:2007sr,Khemchandani:2008rk,MartinezTorres:2008kh,MartinezTorres:2010zv}, thus indicating the importance of a heptaquark  content  in some cases. Proceeding in a similar way we investigate if meson-baryon configurations with a positive strangeness also form a resonance or a bound state. 

By solving coupled channel equations in our formalism with the subtraction constants constrained by the available data, we find no resonance. Further, we attempt to extend our model by considering the exchange of some hyperon resonances in the $u$-channel, for which the necessary couplings are available from our previous works \cite{hyperons}. But we end up finding these contributions to be negligible. As a result, we conclude  that light meson-baryon dynamics does not lead to the formation of any resonance.  Our work, thus, does not support the existence of any light pentaquark with spin-parity $1/2^-$ or $3/2^-$.

\section{The $KN$ and $K^*N$ scattering}\label{intr} 
A study of the $KN$ and $K^*N$ coupled channel dynamics  
requires the calculation of the scattering matrix, $T$. 
In the present work we are interested in obtaining the $T$-matrix in $s$-wave.  This can be done by solving the Bethe-Salpeter equation,  
which in its on-shell factorization form reads as~\cite{osetramos,oller}
\begin{align}
T=(1-VG)^{-1} V,\label{BS}
\end{align}
where $G$ is the loop function of two hadrons. The kernel $V$, or potential, is obtained from the  Lagrangians based on the hidden local symmetry (as done in Refs.~\cite{vbvb,pbvb,hyperons}), when dealing with vector mesons, and on the chiral symmetry, when studying pseudoscalar meson-baryon systems. The transition between $KN$ and $K^*N$  is obtained from an extension of  the Kroll-Ruderman term \cite{pbvb}.

\subsection{Determination of the kernel $V$ of the Bethe-Salpeter equation} \label{potentials}

\subsubsection{Pseudoscalar meson-baryon interaction}\label{pbint} 

To determine the $KN\to KN$ amplitude we use the lowest order chiral Lagrangian \cite{ecker,pich}
\begin{eqnarray}\nonumber
\mathcal{L}_{PB} &=& \langle \bar B i \gamma^\mu \partial_\mu B  + \bar B i \gamma^\mu[ \Gamma_\mu, B] \rangle - M_{B} \langle \bar B B \rangle  \\
&+&  \frac{1}{2} D^\prime \langle \bar B \gamma^\mu \gamma_5 \{ u_\mu, B \} \rangle + \frac{1}{2} F^\prime \langle \bar B \gamma^\mu \gamma_5 [ u_\mu, B ] \rangle,\label{LPB}
\end{eqnarray}
with  $D^\prime = 0.8$, $F^\prime = 0.46$, such that  $F^\prime + D^\prime \simeq  g_A = 1.26$ and 
\begin{equation}\nonumber
\Gamma_\mu = \frac{1}{2} \left( u^\dagger \partial_\mu u + u \partial_\mu u^\dagger  \right), \,u_\mu = i u^\dagger \partial_\mu U u^\dagger ,\,
U=u^2 = exp\left(i \frac{P}{f}\right), 
\end{equation}
where  P and B are SU(3) matrices for pseudoscalar mesons and octet baryons given by
\begin{eqnarray} \nonumber
P =
\left( \begin{array}{ccc}
\pi^0 + \frac{1}{\sqrt{3}}\eta & \sqrt{2}\pi^+ & \sqrt{2}K^{+}\\
&& \\
\sqrt{2}\pi^-& -\pi^0 + \frac{1}{\sqrt{3}}\eta & \sqrt{2}K^{0}\\
&&\\
\sqrt{2}K^{-} &\sqrt{2}\bar{K}^{0} & \frac{-2 }{\sqrt{3}} \eta
\end{array}\right); \hspace{0.5cm}
B =
\left( \begin{array}{ccc}
 \frac{1}{\sqrt{6}} \Lambda + \frac{1}{\sqrt{2}} \Sigma^0& \Sigma^+ & p\\
&& \\
\Sigma^-&\frac{1}{\sqrt{6}} \Lambda- \frac{1}{\sqrt{2}} \Sigma^0 &n\\
&&\\
\Xi^- &\Xi^0 & -\sqrt{\frac{2}{3}} \Lambda 
\end{array}\right).
\end{eqnarray}
The  standard  Weinberg-Tomozawa contribution for the $KN$ system  can be obtained using Eq.~(\ref{LPB}) as  \cite{osetramos}
\begin{align}
V^I_{KN}=-\frac{C^I_{KN}}{4 f_K^2}(\omega+\omega^\prime),\label{VKN}
\end{align}
where the superscript label $I$ indicates the isospin of the meson-baryon system. In the present case we can have total isospin $I=0$ or $1$. The coefficient $C^I_{KN}$ is $0$ ($-2$) for isospin $0$ ($1$), when using average masses for the kaons ($K^0$, $K^+$) and the nucleons  ($n$, $p$). Further, $f_K=113.46$ MeV is the kaon decay constant, and $\omega$ ($\omega^\prime$) corresponds to the energy of the kaon in the initial (final) state.
 
 \subsubsection{Vector meson-baryon interaction}\label{VBintr}
For the case of  $K^* N$ we use the formalism developed in our previous works~\cite{vbvb,hyperons,more} to investigate the interaction of vector mesons with baryons. We review  this formalism  here for the convenience of the reader.  Our formalism lies on the  theory of the  hidden local symmetry (HLS) developed in Ref.~\cite{bando}, which accommodates vector mesons consistently with the chiral symmetry. Let us start the discussion  by writing the SU(2) VB Lagrangian
\begin{equation}
\mathcal{L_{\rm \rho N}} =   \bar{\psi}  i \slashed D  \psi, \label{rhon}
\end{equation}
which has been obtained  through the minimal substitution 
\begin{equation}
\partial_\mu \longrightarrow D_\mu = \partial_\mu + i g \rho_\mu (x),
\end{equation}
and by  requiring that the nucleon fields ($\psi$) transform under the HLS  as 
$\psi \rightarrow h(x) \psi$, where $h (x)$ is an element of the HLS.  Eq.(\ref{rhon}) can be  further extended to
\begin{equation}
\mathcal{L}_{\rho N } = - g \bar{\psi} \left\{ \gamma_\mu \rho^\mu + \frac{\kappa_\rho}{4M} \sigma_{\mu\nu} \rho^{\mu\nu} \right\} \psi, \label{rhoNL}
\end{equation}
recalling the need to reproduce the anomalous magnetic moment of the baryons. It was found in Refs.~\cite{vbvb,more} that the gauge invariance of  Eq.~(\ref{rhoNL}) under the HLS transformation, which requires the invariance of the new term $\bar{\psi} h^\dagger(x)  \sigma_{\mu\nu}  \rho^{\mu\nu} h (x) \psi$,  can be accomplished only when the commutator part of the tensor field $\rho^{\mu\nu}$ is taken into account. This latter term leads to a contact VB interaction (originating from the two vector fields in the commutator term) which gives a large contribution. Thus, we find the tensor part of the VB interaction to be relevant, not only from the point of view of the contribution obtained from the same but also from the point of view of the gauge invariance of the HLS.
The importance of the tensor interaction has also been discussed in other contexts \cite{Hobbs:2013bia,Haidenbauer:2009ad,Shyam:2014dia}. 

The SU(3) generalization of Eq.(\ref{rhoNL}) leads to the Lagrangian
\begin{eqnarray} \label{vbb}
&&\mathcal{L}_{VB}= -g \Biggl\{ \langle \bar{B} \gamma_\mu \left[ V^\mu, B \right] \rangle + \langle \bar{B} \gamma_\mu B \rangle  \langle  V^\mu \rangle  
\Biggr. \\ \nonumber
&&+\left. \frac{1}{4 M} \left( F \langle \bar{B} \sigma_{\mu\nu} \left[ V^{\mu\nu}, B \right] \rangle  + D \langle \bar{B} \sigma_{\mu\nu} \left\{ V^{\mu\nu}, B \right\} \rangle\right)\right\},
\end{eqnarray}
where  $V^{\mu\nu}$ is the  tensor field of the vector mesons,
\begin{equation}
V^{\mu\nu} = \partial^{\mu} V^\nu - \partial^{\nu} V^\mu + ig \left[V^\mu, V^\nu \right]. \label{tensor}
\end{equation}
and $V$ denotes the SU(3) matrix for the vector mesons 
\begin{eqnarray}
V =\frac{1}{2}
\left( \begin{array}{ccc}
\rho^0 + \omega & \sqrt{2}\rho^+ & \sqrt{2}K^{*^+}\\
&& \\
\sqrt{2}\rho^-& -\rho^0 + \omega & \sqrt{2}K^{*^0}\\
&&\\
\sqrt{2}K^{*^-} &\sqrt{2}\bar{K}^{*^0} & \sqrt{2} \phi 
\end{array}\right).
\end{eqnarray}

In Eq.(\ref{vbb}), the coupling $g$ is obtained by the Kawarabayashi-Suzuki-Riazuddin-Fayyazuddin relation~\cite{ksrf1,ksrf2}
\begin{align}
g=\frac{m_{K^*}}{\sqrt{2}f_{K^*}},
\end{align}
and the constants $D$ = 2.4, $F$ = 0.82. These values of $D$ and $F$ were found to well reproduce the magnetic moments of the baryons in Ref.~\cite{jidohosaka}. It should be mentioned here that Eq.~(\ref{vbb}) is also in good agreement with the VB Lagrangians obtained within other approaches  \cite{jenkins, meissner}.

Further, to obtain the right couplings for the physical $\omega$ and $\phi$ meson  at  the meson-baryon-baryon vertices, 
we have considered the mixing of their octet and singlet components. Under the ideal mixing assumption, we write 
\begin{eqnarray}\nonumber
\omega &=& \sqrt{\dfrac{1}{3}} \, \omega_8 + \sqrt{\dfrac{2}{3}} \,\omega_0, \\
\phi &=&  -\sqrt{\dfrac{2}{3}}  \,\phi_8 + \sqrt{\dfrac{1}{3}} \,\phi_0,
\end{eqnarray}
and use only the octet part of these wave function 
in Eq.~(\ref{vbb}). In other words, the Lagrangian given by Eq.~(\ref{vbb})
corresponds to the interaction between the octet vector mesons and the octet baryons. For the singlet states we write
\begin{eqnarray} \label{singlet}
\mathcal{L}_{V_0BB} = -g \Biggl\{  \langle \bar{B} \gamma_\mu B \rangle  \langle  V_0^\mu \rangle  
+ \frac{ C_0}{4 M}  \langle \bar{B} \sigma_{\mu\nu}  V_0^{\mu\nu} B  \rangle  \Biggl\},
\end{eqnarray}
where the constant $C_0$ is chosen to be $3F - D$ such that  the anomalous magnetic 
coupling at the $\phi NN$ vertex is null and for $\omega NN$  is $\kappa_\omega \simeq 3F - D$. These results, together with the anomalous magnetic coupling at the
$\rho NN$ vertex, which is $D + F = \kappa_\rho$, lead  to a consistent formalism.

From Eqs.(\ref{vbb}) and (\ref{singlet}), we can determine Yukawa type vertices which can be used to write diagrams corresponding to $s$- and $u$-channel exchange of a baryon. To obtain the amplitudes for $t$-channel diagrams we additionally need the kinetic term of the hidden local symmetry Lagrangian  for the three-vector meson vertices 
\begin{equation}
\mathcal{L}_{3V} \in - \frac{1}{2} \langle V^{\mu\nu} V_{\mu\nu} \rangle.
\end{equation}

The VB amplitudes in our formalism, hence, get contribution from 
$s$-, $t$-, and $u$-channel exchange diagrams together with a contact term (CT) arising from the commutator in the vector meson tensor. Thus, the leading order amplitude for $K^*N$ is written as
\begin{align}
V^I_{K^* N} = V^I_{t, K^* N} + V^I_{s, K^* N} + V^I_{u, K^* N} + V^I_{\textrm{CT}, K^* N}.\label{VKsN}
\end{align}
Since $K^*$ has spin $1$ and the spin of the nucleon $N$ is $1/2$, we can have total spin $S=1/2$ and $3/2$ for the system in $s$-wave. The potentials in Eq.~(\ref{VKsN}) are, in fact,  spin- (and isospin-) dependent and need to be projected to each configuration. The $t$-channel amplitude in Eq.~(\ref{VKsN}) is analogous to the one in Eq.~(\ref{VKN}) (as also obtained earlier in Ref.~\cite{ramosvb}),
\begin{align}
V^I_{t, K^* N}=-\frac{C^I_{t, K^* N}}{4 f_{K^*}^2}(\omega + \omega^\prime)  \vec{\epsilon}_1\cdot \vec{\epsilon}_2,\label{VtK*N}
\end{align}
with $\omega (\omega^\prime)$ and $\epsilon_1 (\epsilon_2)$ representing the energy and the polarization vector of the $K^*$ in the initial (final) state, respectively. The values of  $C^I_{t, K^* N}$ are  $0$ and $-2$ for isospin $0$ and $1$, respectively, and $f_{K^*} = 171.12$ MeV is the decay constant of $K^*$ \cite{Maris:1999nt,Maris:1999sz}. 

Next, for the system studied here, the $s$-channel potential is trivially zero, since it would imply the exchange of a baryon with strangeness $+1$. The $u$-channel and the contact term are given, at nonrelativistic energies, by
\begin{align}
V^I_{u, K^*N}&=C^I_{u, K^*N} \left(\frac{g^2}{2\bar M - m}\right) \vec{\epsilon}_1\cdot \vec{\sigma}\,\, \vec{\epsilon}_2\cdot \vec{\sigma },\nonumber\\
V^I_{\textrm{CT}, K^*N} &=i C^I_{\textrm{CT}, K^*N}\,\frac{g^2}{2 M}\vec{\sigma}\cdot\vec{\epsilon_2} \times \vec{\epsilon_1},\label{Vuc1}
\end{align}
where $m$ (M)  is the mass of the $K^*$ (N) and $\bar M$ represents an average mass for the baryons involved in the process. The coefficients $C^I_{u, K^*N}$ for the exchange of an octet baryon and $C^I_{\textrm{CT}, K^*N}$ are given in Table~\ref{T1}. It is interesting to notice that  $V_{t, K^*N}$ alone is spin degenerate, while the total potential in Eq.~(\ref{VKsN}) is spin-isospin dependent due to the structure of the amplitudes coming from the $u$-channel and the contact term (given in Eq.~(\ref{Vuc1})). This finding shows that  diagrams other than $t$-channel can play an important role in studies related to VB systems.
\begin{table}[h]
\caption{Isospin coefficients for the $u$-channel amplitude given by Eq.~(\ref{Vuc1}). 
}\label{T1}
\centering
\begin{tabular}{c|c|c}
\hline\hline
&$I=0$&$I=1$\\
\hline
&&\\
$C^I_u$&$\frac{D m [(D-3F) m-6\bar M]}{6 \bar M^2}$&$\frac{12 \bar M^2+12 F m \bar M +(D^2+3F^2) m^2}{12 \bar M^2}$\\
&&\\
$C^I_\textrm{CT}$& $D$&$-F$\\
\hline\hline
\end{tabular}
\end{table}

\begin{figure}[h!]
\begin{center}
\includegraphics[width=0.9\textwidth]{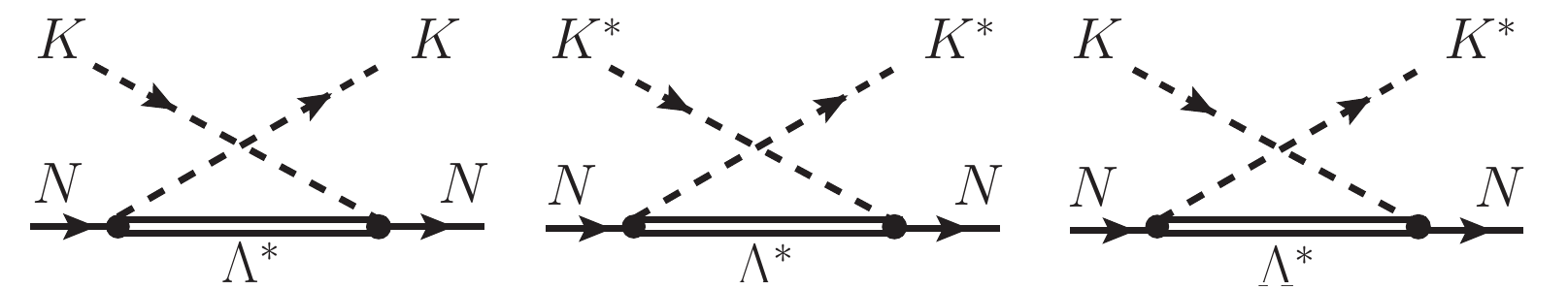}
\caption{Diagrams involving $u$-channel exchange of $\Lambda$ resonances.}\label{uexcL}
\end{center}
\end{figure}

It is possible to argue that, in principle, exchange of a baryon resonance with negative parity and/or higher spin may also contribute to the $u$-channel amplitude. We have not considered such a possibility in our previous works since the couplings of several resonances to different meson-baryon channels are not well known and, hence, it can lead to the introduction of  numerous unknown parameters in the  formalism which can be difficult to control. However, in the present work, we can consider at least the exchange of the resonances found in Ref.~\cite{hyperons}, where the same formalism was applied to strangeness $-1$ systems.  The states found in Ref.~\cite{hyperons}  can be associated with some well known resonances: $\Lambda(1405)$, $\Lambda(1670)$, $\Lambda(2000)$, $\Sigma(1750)$, $\Sigma(1940)$ and $\Sigma(2000)$. Thus, we  consider the exchange  of these states in the $u$-channel (see examples of such diagrams in Fig.~\ref{uexcL}) by using the couplings  obtained for both PB and VB channels in Ref.~\cite{hyperons}.

To determine the contribution from the exchange of resonances, we write the following phenomenological effective field Lagrangians \cite{nam,ozaki}
\begin{align}
\mathcal{L}_{NK H^*}&=ig_{NK H^*}\bar{H}^* K^\dagger N+\textrm{h.c},\nonumber\\
\mathcal{L}_{NK^* H^*}&=g_{NK^*H^*}\bar{H}^* \gamma^\mu\gamma_5 K^{*\dagger}_\mu N + \textrm{h.c},\label{L2}
\end{align}
where $H^*$ stands for $\Lambda^*$ or $\Sigma^*$ and $K^\dagger (K^{*\dagger})$  creates a $K$ ($K^*$) meson.
 It should be remarked here that 
the couplings $g_{NK\Lambda^*}$ and $g_{NK^*\Lambda^*}$ were obtained in Ref.~\cite{hyperons} by studying VB interaction using the Lagrangian given in  Eq.~(\ref{vbb}) and, thus, the contributions from both vector and tensor VB interaction are embedded in  $g_{NK\Lambda^*}$ and $g_{NK^*\Lambda^*}$ in Eq.~(\ref{L2}).
We list $g_{NK\Lambda^*}$ and $g_{NK^*\Lambda^*}$ in Table~\ref{couplings}  for completeness.
\begin{table}[h!]
\centering
\caption{Coupling of the resonances considered in the $u$-channel diagram with different meson baryon channels, as found in Ref.~\cite{hyperons}.
There are two values listed for $\Lambda (1405)$ and $\Sigma(2000)$ in this table since a two pole structure was found to be associated to them in Ref.~\cite{hyperons}. 
} \label{couplings}
\begin{tabular}{c|c|c}
\hline
Resonance ($R$)& $g_{KNR}$&$g_{K^*NR}$\\
\hline
$\Sigma(1940)$ $D_{13}$&$0.0 + i 0.0$&$-0.3+i0.2$\\
\hline
$\Lambda(2000)$ $S_{01}$&$-0.2+i0.4$&$-1.1+i0.9$\\
\hline
$\Sigma(1750)$ $S_{11}$&$0.1+i0.4$&$2.7+i1.2$\\
\hline
\multirow{2}{*}{$\Sigma(2000)$ $S_{11}$}&$0.9-i0.6$&$-0.8-i0.1$\\
&$-0.2-i0.4$&$0.9+i0.1$\\
\hline
\multirow{2}{*}{$\Lambda(1405)$ $S_{01}$}&$1.2-i1.4$&$0.4+i1.6$\\
&$2.8+i0.6$&$-4.9-i0.0$\\
\hline
$\Lambda(1670)$ $S_{01}$&$0.3-i0.6$&$0.9+i0.3$\\
\hline
\end{tabular}
\end{table}

Using the Lagrangians of Eqs.~(\ref{L2}), we obtain the diagonal amplitudes for the diagrams shown in Fig.~\ref{uexcL}, within the non relativistic approximation (consistent  with the procedure followed to obtain the  amplitudes in Eqs.~(\ref{VtK*N}) and (\ref{Vuc1})), as
\begin{align}
V^{H^*}_{u,K}&= D_I \mid g_{NKH^*} \mid^2 \frac{M-m_K+M_{H^*}}{u-M^2_{H^*}+ i M_{H^*} \Gamma_{H^*}},\label{ampudia1}\\
V^{H^*}_{u,K^*}&=D_I \mid g_{NK^*H^*} \mid^2 \frac{M-m_{K^*}+M_{H^*}}{u-M^2_{H^*}+ i M_{H^*} \Gamma_{H^*}} \vec{\epsilon_1}\cdot\vec{\sigma}\,\vec{\epsilon_2}\cdot\vec{\sigma},\label{ampudia2}
\end{align}
where $u$ is the Mandelstam variable, $M_{H^*}$ and $\Gamma_{H^*}$ are the mass and the width of the exchanged $\Lambda^*$ or $\Sigma^*$ resonance. 
 $D_I $  in Eqs.~(\ref{ampudia1},\ref{ampudia2}) is  a Clebsh-Gordon coefficient  taking care of the fact that we use  $g_{NK\Lambda^*}$ and $g_{NK^*\Lambda^*}$ from Ref.~\cite{hyperons} which are isospin projected couplings
 while the $\Lambda$'s can be exchanged only in the processes $K^0 p \leftrightarrow K^+ n$ and $K^{*0} p \leftrightarrow K^{*+} n$. For these processes, the value of $D_I$ is $-1$  for a $\Sigma^*$ exchange and $-1/2$ in case of  $\Lambda^*$. For diagonal processes proceeding through a $\Sigma^*$ exchange $D_I$ is 1.

\subsubsection{Amplitudes for PB $\leftrightarrow$ VB} \label{pbvbint}
The transition $KN \to K^* N$ amplitude is obtained from the PBVB Lagrangian deduced in Ref.~\cite{pbvb}
consistently with the HLS. The procedure consists  of  using the Kroll-Ruderman term for the photoproduction of a pion and replacing
the photon  by a vector meson.  To  do this we use the $\pi N$ Lagrangian from the non-linear sigma model 
\begin{equation}
\mathcal{L}_{\pi N} = \bar{\psi} \left[ i \gamma^\mu \partial_\mu - g_{\pi NN} \left( \sigma + i \vec{\tau}.\vec{\pi} \gamma_5 \right)  \right] \psi,\label{lpin}
\end{equation}
and 
introduce a vector meson field as a gauge boson of the HLS through:
$i \slashed \partial \longrightarrow i \slashed\partial - g \slashed \rho$,
to obtain
\begin{eqnarray}
\mathcal{L}_{\pi N \rho N} &=& 
 - i \frac{g g_A}{2 f_\pi} \bar{N}   \left[ \pi ,  \rho^\mu \right]  \gamma_\mu \gamma_5 N,\label{lpnrn}
\end{eqnarray}
where   $\pi = \vec{\tau} \cdot \pi$ and $\rho = \vec{\tau} \cdot \dfrac{\rho}{2}$.

Generalizing the Lagrangian in Eq.~(\ref{lpnrn}) for SU(3) leads to
\begin{equation}
\mathcal{L}_{\rm PBVB} = \frac{-i g_{KR}}{2 f_\pi} \left ( F^\prime \langle \bar{B} \gamma_\mu \gamma_5 \left[ \left[ P, V^\mu \right], B \right] \rangle + 
D^\prime \langle \bar{B} \gamma_\mu \gamma_5 \left\{ \left[ P, V^\mu \right], B \right\}  \rangle \right), \label{pbvb}
\end{equation}
where  $F^\prime = 0.46$, $D^\prime = 0.8$ such that  $F^\prime + D^\prime \simeq  g_A = 1.26$ \cite{pbvb}.

The $KN \to K^* N$ amplitude obtained using Eq.~(\ref{pbvb}) is
\begin{align}
V^I_{KN  K^*N}=i\sqrt{3}\frac{g_{\textrm{KR}}}{2\sqrt{f_K f_{K^*}}} C^I_{KN K^*N},\label{KR}
\end{align}
where $g_{\textrm{KR}}$ is the Kroll-Ruderman coupling \cite{pbvb}
\begin{align}
g_{\textrm{KR}}=\frac{m_{K^*}}{ \sqrt{2 f_K f_{K^*}}}\sim 4.53. \label{gkr}
\end{align}
The isospin coefficient $C^I_{KN K^*N}$ is  $-2D^\prime$ for isospin 0 and $2F^\prime$ for isospin 1.


Note that in our formalism  PB and VB channels couple only in the spin $1/2$ configuration. Thus, the
amplitude in Eq.~(\ref{KR}) determines the transition $KN \to K^*N$ for isospins $0, 1$ and  total spin $S=1/2$. The PB-VB coupling 
for total spin 3/2 is zero (as in Ref.~\cite{pbvb,hyperons,more}). This is consistent with the results obtained within a different formalism ~\cite{javi}, where  the  VB amplitudes in spin 3/2 have been found to change weakly when coupled  to pseudoscalar baryon  systems.

As can be seen in Fig.~\ref{uexcL}, the transition amplitude $KN\to K^*N$ can also get a contribution arising from the $u$-channel exchange of hyperon resonances. 
Using the Lagrangians given in Eq.~(\ref{L2}), we obtain
\begin{equation}
V^{H^*}_{KN K^*N}= i D_I \mid g_{NKH^*} \mid  \mid g_{NK^*H^*} \mid \frac{M-m_{K^*}+M_{H^*}}{u-M^2_{H^*}+ i M_{H^*} \Gamma_{H^*}} \vec{\epsilon}\cdot\vec{\sigma},
\end{equation}
where $H^*$, $\Gamma^*$ and $D_I$ have same meaning and values as in Eqs.~(\ref{ampudia1}, \ref{ampudia2}).

Finally,  we must mention that the following form factor is multiplied to $u$-channel amplitudes, following Refs.~\cite{vbvb,hyperons},
\begin{equation}
F(\Lambda, u) =\frac{\Lambda^4}{\Lambda^4 + (u - M_u^2)^2},\label{formf} 
\end{equation}
where $u$ is the usual Mandelstam variable, $M_u$ is the  mass of the baryon exchanged and $\Lambda$ is a cut-off which 
we vary in the range 650-1000 MeV. 
As discussed in Refs.~\cite{vbvb,hyperons}, only the terms involving the negative energy solution of the Dirac equation for the baryon propagator contribute for such diagrams when studying  near threshold $s$-wave meson-baryon interaction. The above form factor takes care of the fact that such diagrams require large momentum transfers at the non-relativistic energies.

\subsection{The loop function}
After determining the kernel $V$ needed to solve the Bethe-Salpeter equation, the other element required to obtain the scattering matrix is the loop function $G$ of Eq.~(\ref{BS}),
which is given by
\begin{align}\nonumber
G (\sqrt{s}, m, M) = i \,2 M \int \dfrac{d^4q}{(2\pi)^4} \frac{1}{(P - q)^2 - M^2 + i\epsilon} \frac{1}{q^2 - m^2 + i\epsilon},
\end{align}
where $\sqrt{s}$ is the center of mass energy, $M$ ($m$) corresponds to the mass of the nucleon (meson) present in the loop and $P$ is the total four momenta
of the system. As can be seen, this loop function is logarithmically divergent and it needs to be regularized. Standard procedures to calculate the loop functions involve using a three-momentum cut-off or dimensional regularization. In this paper, we use the latter scheme,
in which case, the loop function is written, in the center of mass frame (CM), as~\cite{jido,Gamermann:2010zz,vbvb,more,carmen2,GarciaRecio:2005hy,javi,oller,ramosvb}
\begin{eqnarray}\label{loop}
G (\sqrt{s}, m, M) &=&\dfrac{2M}{16\pi^2} \Biggl\{ b (\mu) + \ln \dfrac{M^2}{\mu^2} + \dfrac{m^2-M^2+s}{2s}\ln \dfrac{m^2}{M^2} \Biggr.\\
&&+\dfrac{\bf {q}}{\sqrt{s}} \Bigl[ \ln\left(s-\left( M^2-m^2 \right) + 2\bf {q}\sqrt{s}\right)\Bigr.+\ln\left(s+ \left( M^2 - m^2 \right)+2\bf {q}\sqrt{s}\right)\nonumber\\
&&-\ln\left(-s +\left( M^2 - m^2 \right) + 2\bf {q}\sqrt{s}\right)-\Biggl. \Bigl. \ln\left(s- \left( M^2 - m^2 \right) + 2\bf {q}\sqrt{s}\right) \Bigr] \Biggr\}. \nonumber
 \end{eqnarray}
In Eq.~(\ref{loop}), $\bf {q}$ is the on-shell momentum of the particles in the CM, $\mu$ is the regularization scale and $b(\mu)$ a subtraction constant, which needs to be fixed, normally, by requiring the amplitudes to fit some experimental data. Thus, the only parameters to be fixed are the subtraction constants required to regularize the loops (since any change in $\mu$ can be reabsorbed in the value of the subtraction constant $b(\mu)$).

\subsection{Calculation of phase shifts and scattering lengths}
With these ingredients we solve Eq.~(\ref{BS}) and obtain the scattering matrix for the $KN$-$K^*N$ system. The subtraction constants $b_{KN}$ and $b_{K^* N}$ present in the loop functions of $KN$ and $K^* N$ are fixed by fitting the data available from the partial wave analysis groups \cite{prc75,gwu,prc29} on the isospin $0$ and $1$ $KN$ phase shifts ($\delta^0_{KN}$ and $\delta^1_{KN}$, respectively), which, in our formalism, are related to the scattering matrix through the relations~\cite{MartinezTorres:2008kh}
\begin{equation}
\widetilde{T}^{I,S} = \left(\frac{\eta^{I,S}\, e^{2i \delta^{I,S}}-1}{2 i }\right),\label{delta}
\end{equation}
with
\begin{align}
T^{I,S}=-\frac{4\pi\sqrt{s}}{M \bf {q}} ~~\widetilde{T}^{I,S}, 
\end{align}
where $\bf {q}$ is the center-of-mass momentum and $\eta^{I,S}$ is the inelasticity in the isospin $I$ and spin $S$. Finally, using the resulting scattering matrix we also calculate the $KN$ and $K^*N$ scattering lengths for different isospin and spin, $a^{I,S=1/2}_{KN}$ and $a^{I,S}_{K^* N}$, using the relation~\cite{osetramos}
\begin{align}\label{scatl}
a^{I,S} = -\frac{M}{4\pi \sqrt{s}}T^{I,S},
\end{align}
at  threshold energies.

\section{Results and discussions}\label{results}

Let us start the discussions on the results by recalling that the Weinberg-Tomozawa interaction for the $KN$  channel  is null for isospin~0 (considering isospin averaged masses in Eq.~(\ref{VKN}), which leads to $C^I_{KN}=0$), which gives null scattering phase shifts. Even the consideration of the differences between the masses leads to nearly zero $KN$ potential, as shown in Ref.~\cite{osetramos} where a scattering length of the order of $10^{-7}$ fm was obtained.  While a nearly zero scattering length is  compatible with the results of some older partial wave analyses \cite{KNold}, some other report values varying between $-0.1$ fm \cite{prc75} (which is the most recent one) to $-0.4$ fm \cite{martin}.   To reproduce these values and the available data on the $S_{01}$ (representing $L_{\rm Isospin,\,\, 2\times Spin}$) partial wave $KN$ phase shifts \cite{prc75,gwu,prc29}, it is required to go beyond the $KN$  interaction obtained from the lowest order chiral Lagrangian. One possibility is to consider contributions from higher order terms of the Lagrangian (like Refs.~\cite{Oller:2005ig, Guo:2015xva}). While that would imply fixing a large number of parameters with scarce data available, it is wiser and important to exhaust possible corrections to be included while keeping the lowest order contribution for the $KN$ system. For example, including the coupling to the $K^* N$ channel, which is the aim of the present paper.  

As discussed in the previous section, we consider $t$- , $u$-channel diagrams and a contact interaction to obtain the $K^* N$ diagonal amplitudes.  The $t$-channel interaction in isospin 0 is found to give a null amplitude (as also in Ref.~\cite{GarciaRecio:2005hy}), while the contribution from other diagrams: contact term and the  $u$-channel diagrams lead to nonzero contributions.  Out of these different contributions, we find that the ones arising from the exchange of hyperon resonances in the $u$-channel diagrams (shown in Fig.~\ref{uexcL}) are negligibly small in the present case. 
Such contributions have been found to be important in the studies of some processes involving nonstrange and strangeness $-1$ systems~\cite{Shyam:2011ys,Ozaki:2009mj}.  However,  we do not find this to be the case for $KN$-$K^*N$ systems. 
Indeed the exchange of the octet baryons was also found to give important contribution to some processes and negligible to others in Refs.~\cite{vbvb,hyperons,more}.

\begin{figure}[h!]
\centering
\includegraphics[width=0.45\textwidth]{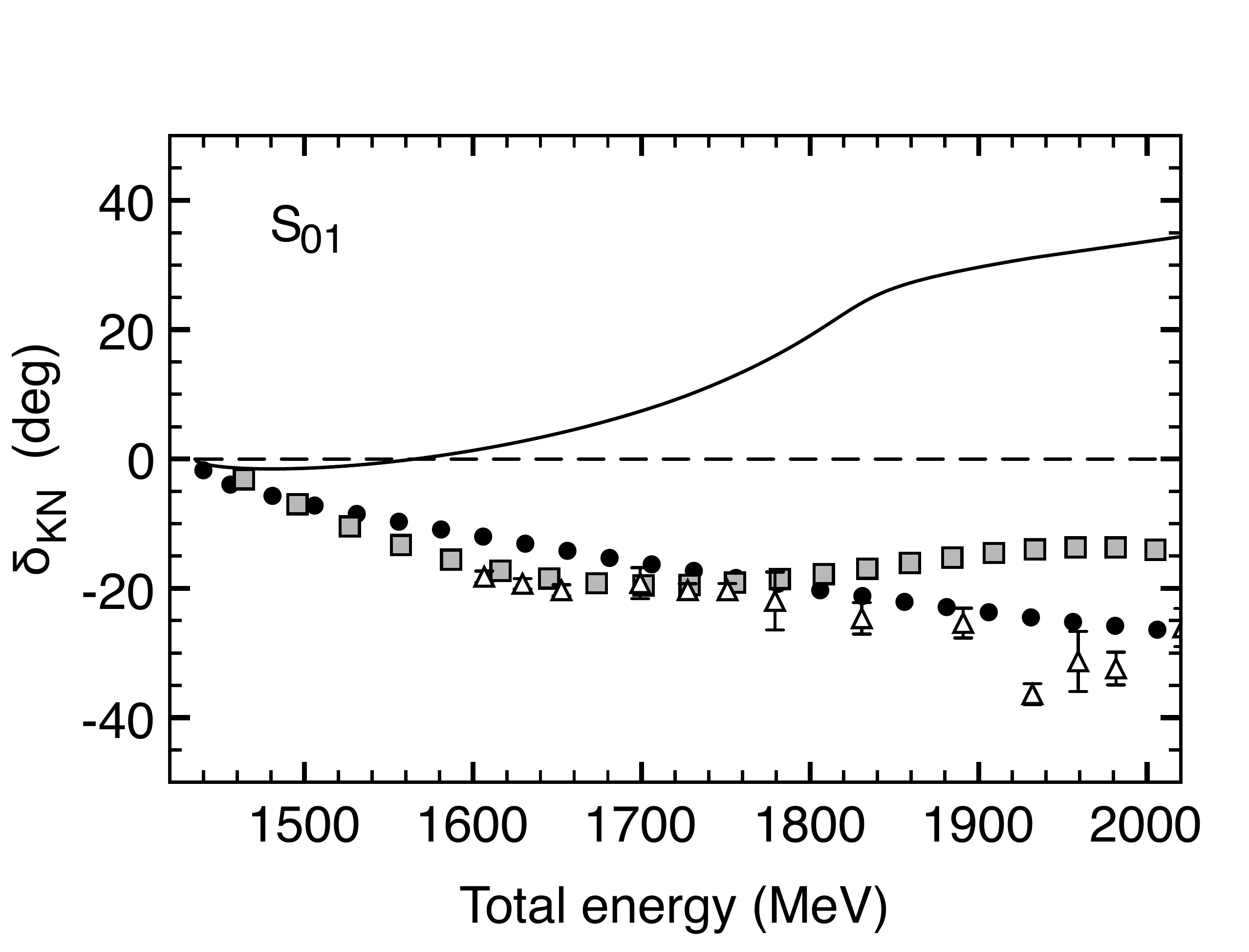}
\includegraphics[width=0.45\textwidth]{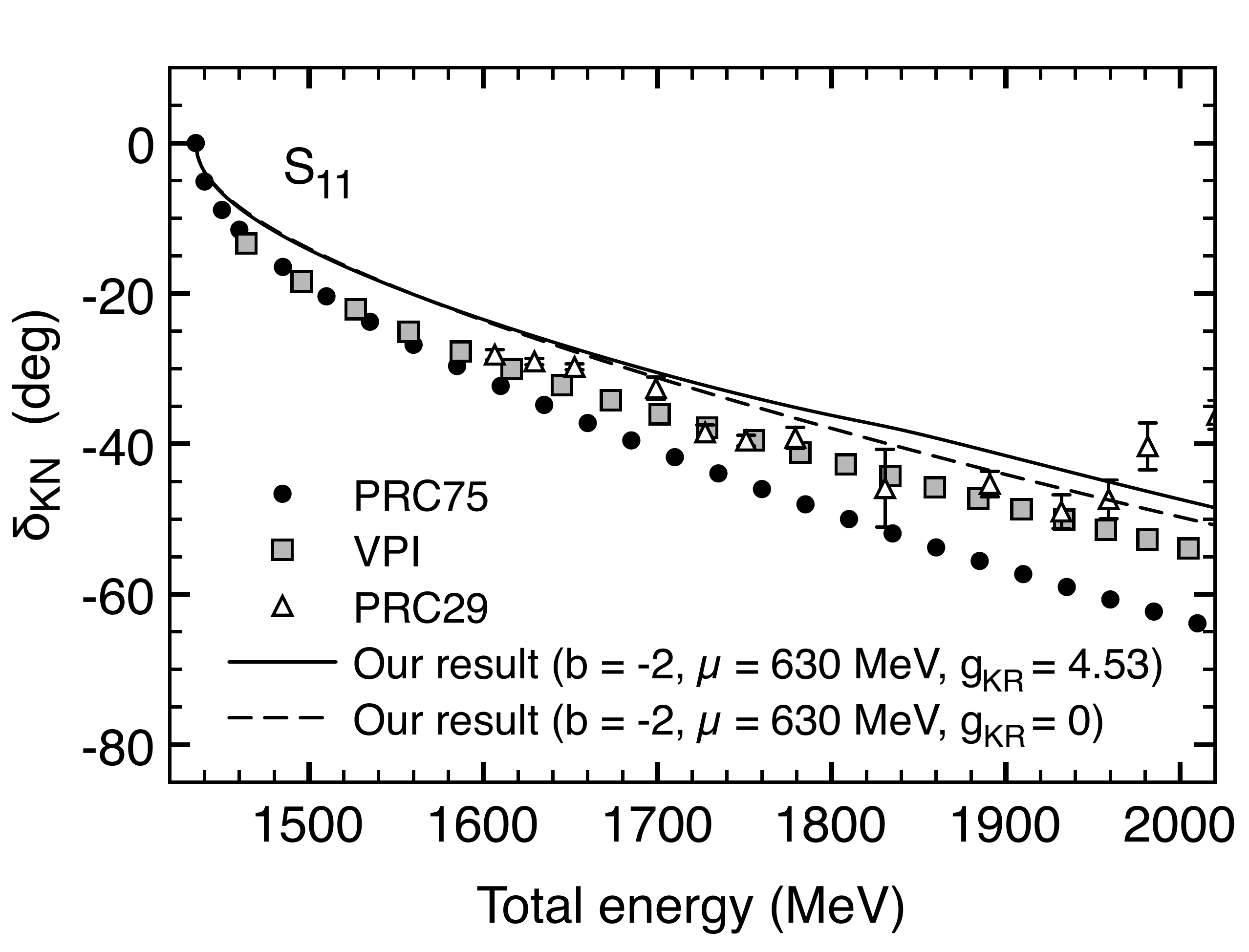}
\caption{Scattering phase shifts for the $KN$ system in $S_{01}$ and $S_{11}$ partial waves. The solid lines show the results obtained within our $KN$-$K^*N$ coupled channel  formalism (i.e., using the value of $g_{\rm KR}$ given by Eq.~(\ref{gkr})) by using the same subtraction constants as those used to reproduce the experimental data for the strangeness $-1$ coupled channel scattering \cite{osetramos}. The dashed lines are obtained using the same subtraction constants but setting the coupling between $KN$-$K^*N$ to zero. The data from the partial wave analysis, represented by filled circles, boxes and empty triangles are taken from Refs.~\cite{prc75,gwu,prc29}, respectively. }\label{one}
\end{figure}

To solve the Bethe-Salpeter  equation, we need to fix the subtraction constants which are required to regularize the logarithmical divergence of the loop functions. One way to proceed with the calculations would be to consider the same ``natural" values \cite{hyodo} of the subtraction constants ($b(\mu) = -2$, with $\mu$ =630 MeV) as the ones used for the meson-baryon systems with the opposite strangeness  (S = $-1$) \cite{osetramos}. Let us denote these subtraction constants as  the ``Parameter set I" in order to simplify the subsequent discussions. In Fig.~\ref{one} we show the isospin 0 (left panel) and 1 (right panel) $KN$ phase shifts obtained with these constants for: (a) the coupling  between $KN$ and $K^*N$ given by Eq.~(\ref{gkr}) (as solid lines) and (b) for its value $g_{\rm KR} = 0$ (as dashed lines).

As can be seen from this figure, the isospin 1 phase shifts data~\cite{prc75,gwu,prc29} (shown in the right panel) can be reasonably reproduced by considering $K N$ channel alone and by using the parameter set I (dashed line). The phase shifts in isospin 0 are zero in this case. The zero phase shift in isospin 0 comes trivially from $V_{KN}^{I=0} = 0$   (as explained in section \ref{pbint}). Further,  Fig.~\ref{one} shows that the coupling to the $K^* N$ channel  does not alter much the isospin 1 results but the isospin 0 phase shifts do get affected, although the results do not agree with the data (except near the threshold).
The fact that results for isospin 0 are sensitive to the coupling to the $K^*N$ channel while the isospin 1 are not can be understood by looking at the leading order amplitudes  ($V^I_{t, K^*N}$, $V^I_{\textrm{CT}, K^* N}$, $V^I_{u, K^*N}$, $V^I_{KN  K^*N}$) obtained in sections \ref{VBintr} and \ref{pbvbint}. The sum of $K^*N$ amplitudes (Eqs.~(\ref{VtK*N},\ref{Vuc1})) and the transition $KN \leftrightarrow K^*N$  (see Eq.~(\ref{KR})) all turn out to be very weak in the isospin 1, unlike the case of isospin 0. In other words, the $KN$ channel is found to couple weakly with a weak $K^* N$ amplitude in the isospin 1 configuration.

The poor agreement  of the results obtained in the isospin 0 case suggests that we need to use some other values of the subtraction constants. In principle, the physics related to the strangeness $-1$ and $+1$ meson-baryon systems is different since the presence of a  $s$- or a $\bar s$-valence quark leads to very different situations: the former allows for the existence of a three quark intermediate state while the latter does not. Thus, the subtraction constants required to regularize the loops do not need to be necessarily the same in the two cases. We, thus, make $\chi^2$-fits to the data~\cite{prc75,gwu,prc29} on the $s$-wave isospin 0 and 1 $KN$ phase shifts to fix the subtraction constants. We treat the subtraction constants as free parameters, with the motivation to obtain $KN$ amplitudes in agreement with the data and, hence, obtain constrained predictions for the $K^*N$ amplitudes. 

\begin{figure}[h!]
\centering
\includegraphics[width=0.45\textwidth]{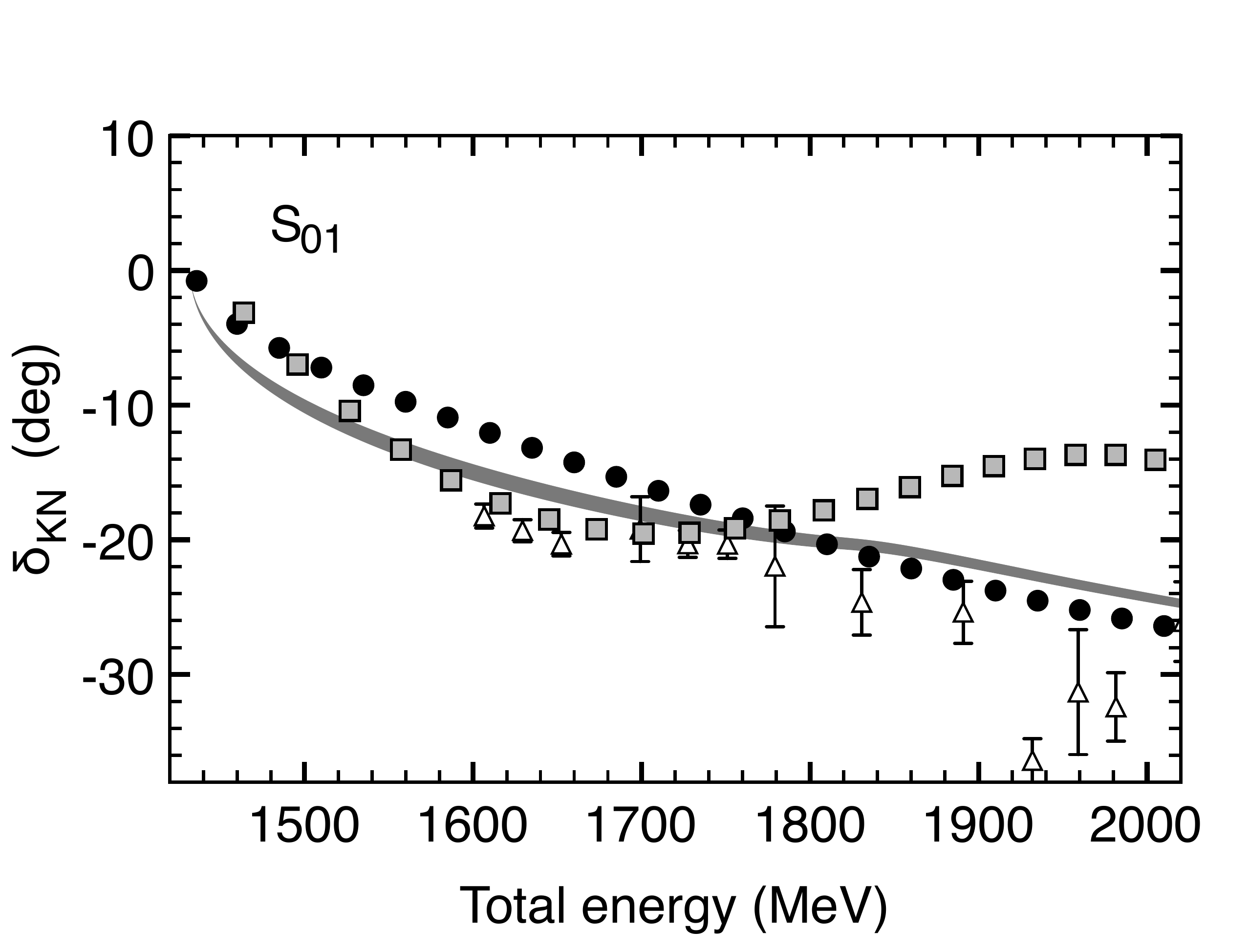}
\includegraphics[width=0.45\textwidth]{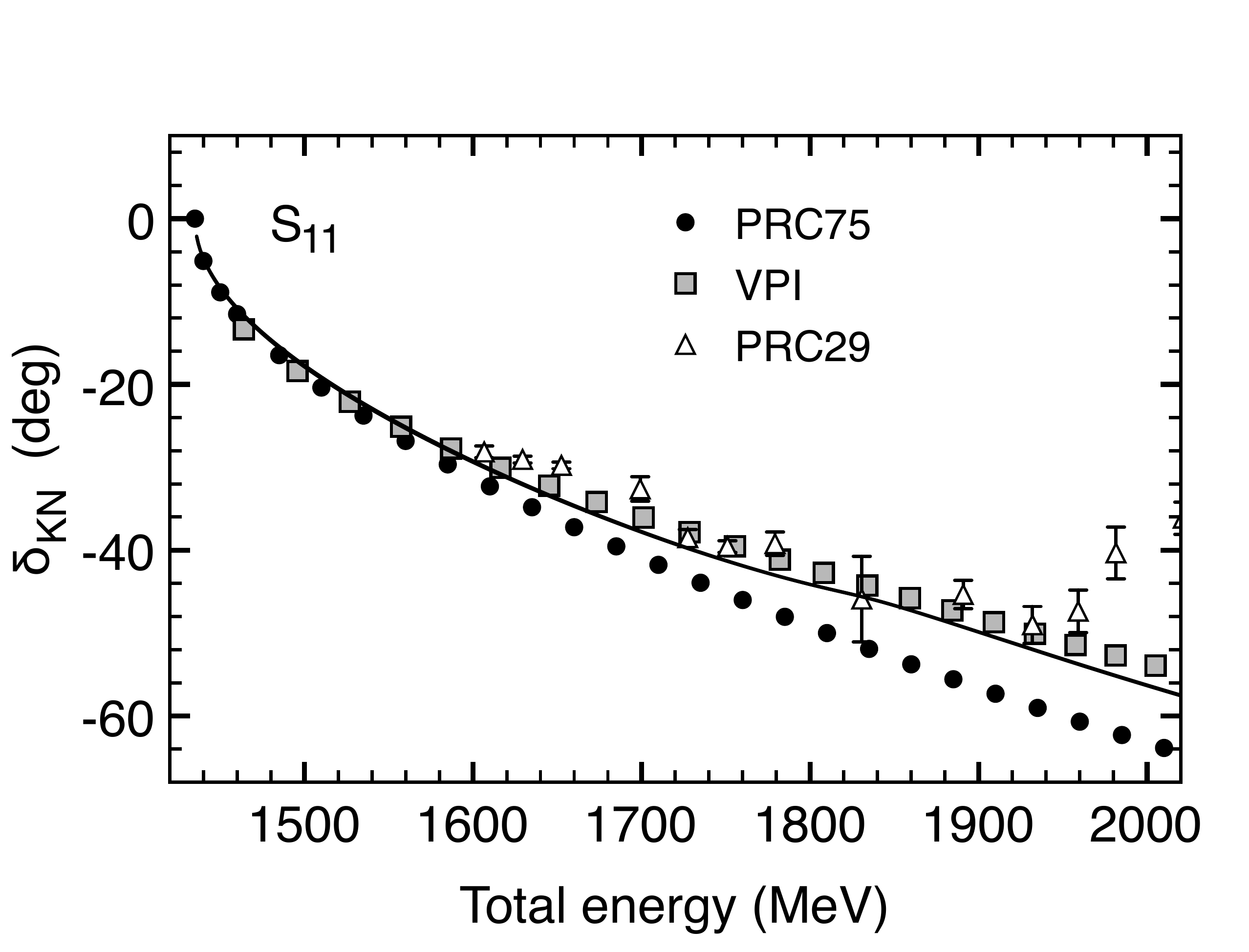}
\caption{Scattering phase shifts for the $KN$ system obtained  by using the parameter set II (subtraction constants: $b^{I=0}_{KN}=-6.82$, $b^{I=0}_{K^*N}=1.84$, $b^{I=1}_{KN}=-1.59$ and $b^{I=1}_{K^*N}=-1$, with $\mu=630$ MeV), and for $\Lambda$ in Eq.~(\ref{formf}) varying between 650-1000 MeV. The lines and symbols here have the same meaning as in Fig.~\ref{one}. }\label{two}
\end{figure}

We find that a good fit is obtained with the subtraction constants: $b^{I=0}_{KN}=-6.82$, $b^{I=0}_{K^*N}=1.84$, $b^{I=1}_{KN}=-1.59$ and $b^{I=1}_{K^*N}=-1$ for the regularization scale $\mu$ fixed to 630 MeV. We shall refer to these values collectively as ``Parameter set II". The resulting phase shifts are shown in Fig.~\ref{two}, 
which also displays the sensitivity of our results to the cut-off parameter $\Lambda$ in Eq.~(\ref{formf}) varied in the range 650-1000 MeV.  It can be seen that the data are reasonably reproduced and that our results are quite stable against the variation in $\Lambda$. We shall, thus, keep the value of $\Lambda =$ 650 MeV for showing further results. 

It can be noticed that the phase shifts shown in Fig.~\ref{two} for isospin $0$ appear like a narrow band while the results for isospin $1$ appear like a line. Also, the results for isospin 0 shown in Figs.~\ref{one} and \ref{two} differ much more than those for isospin 1. This occurs because (1) the subtractions constants are more different in the isospin 0 case, and (2) due to a  stronger influence of $KN$-$K^*N$ coupling and $K^*N$ amplitudes in the isospin 0 configuration, which is weaker in isospin 1.  

Although a good fit to data has been obtained as shown in Fig.~\ref{two}, we must add that the subtraction constants required to fit the data are isospin dependent and are far from natural values, especially in the case of isospin 0. As explained in Refs.~\cite{hyodo,more}, a deviation of subtraction constants from the natural value ($b \simeq -2$ in the present case) can be interpreted as a modification of the interaction kernel, and this modification can be spin-isospin dependent. For example, in a single channel case, we can write
\begin{equation}
T^{I,S} =  \frac{1}{\left(V^{I,S}\right)^{-1} -  G^{I,S}_{\rm phen}},\label{tpheno}
\end{equation}
where $G^{I,S}_{\rm phen}$ is the loop function obtained with the ``phenomenological" subtraction constants which fit the data. Let us denote the loop function obtained using natural values of the subtraction constants by $G_{\rm nat}$ (which is spin, isospin independent) and let $\Delta a$ be a (constant) such that
\begin{equation}
 G^{I,S}_{\rm phen} = G_{\rm nat}+ \Delta a^{I,S} 
\end{equation}

We can now rewrite Eq.~(\ref{tpheno}) in terms of $G_{\rm nat}$ as
\begin{equation}
T^{I,S} = \frac{1}{\left(V^{I,S}\right)^{-1} - G_{\rm nat} -\Delta a^{I,S}} = \frac{1}{\left[ \left(V^{I,S}\right)^{-1} -\Delta a^{I,S}\right]- G_{\rm nat} },
\end{equation}
where the inverse of the expression in the rectangular bracket can be considered as a modified kernel
\begin{equation}
V_{\rm modified}^{I,S} =  \left[ \left(V^{I,S}\right)^{-1} -\Delta a^{I,S}\right] ^{-1}.\label{vmodified}
\end{equation}

\begin{figure}[h!]
\centering
\includegraphics[width=0.43\textwidth]{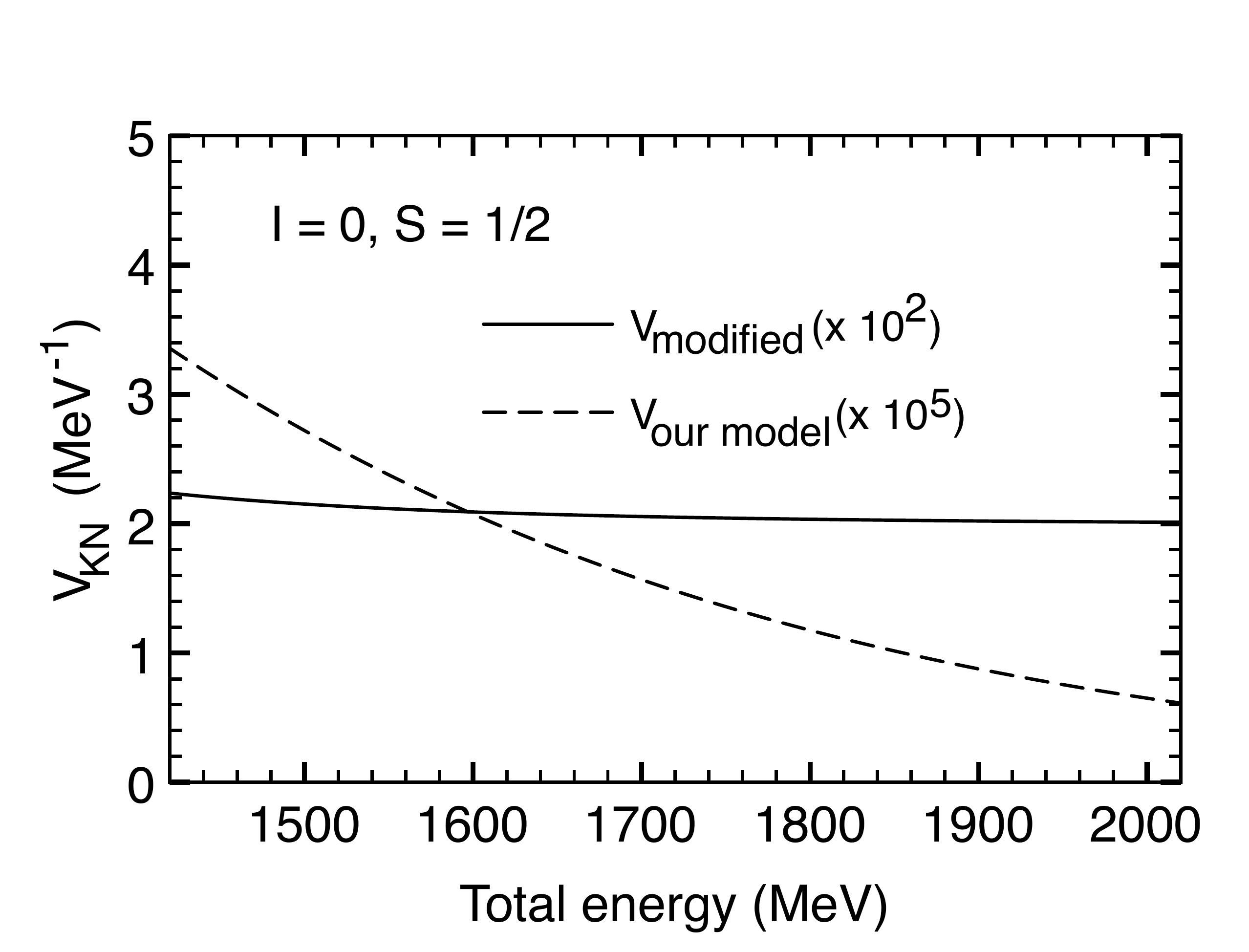}
\includegraphics[width=0.43\textwidth]{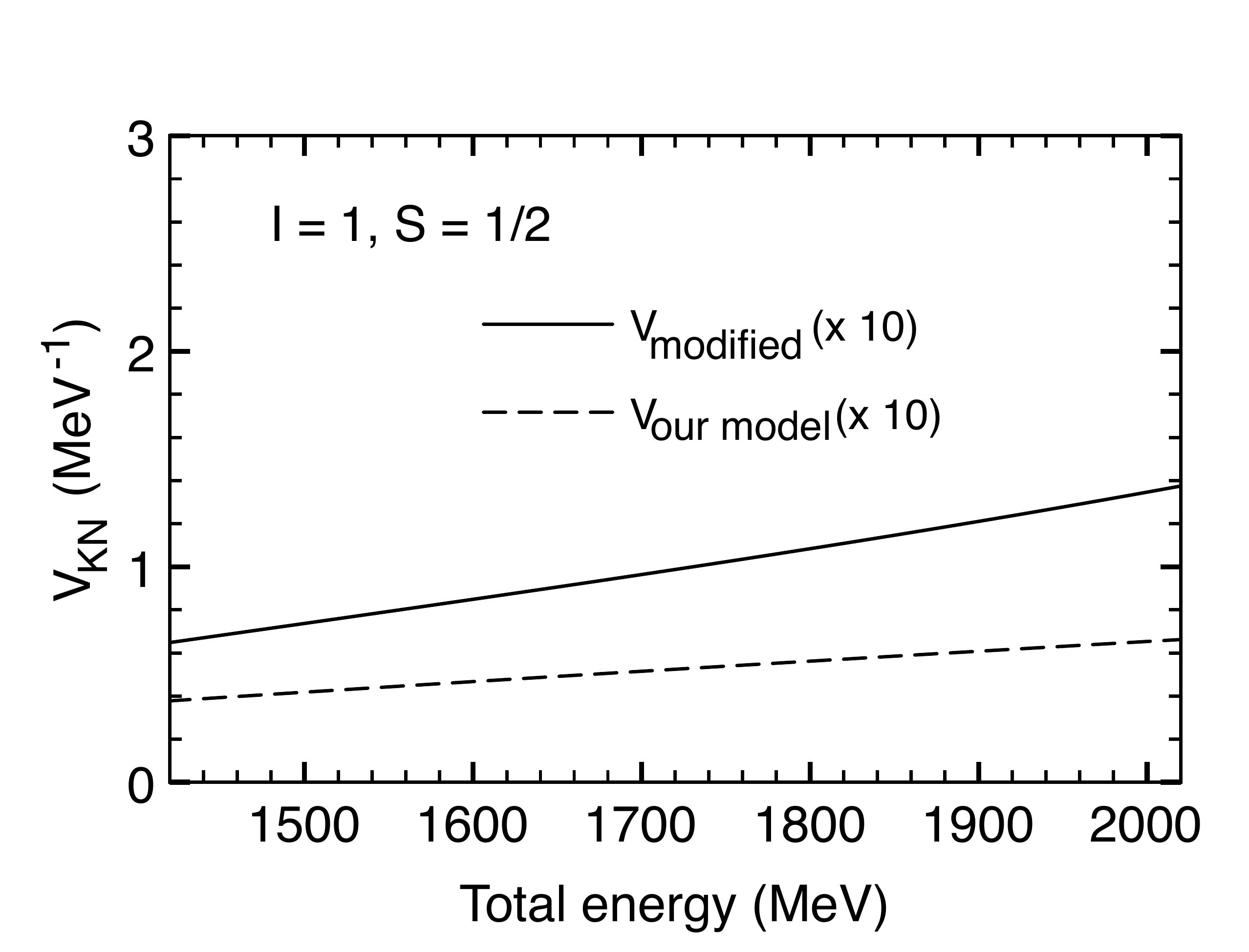}
\caption{A comparison of the kernel $V$ for $KN$ system obtained in Section~\ref{potentials} and defined by Eq.~(\ref{vmodified}).  Here $V_{\rm our~model}$ is the sum of the amplitudes obtained from Eqs.~(\ref{VKN})  and (\ref{ampudia1}). }\label{fig:VKN}
\end{figure}

\begin{figure}[h!]
\centering
\includegraphics[width=0.43\textwidth]{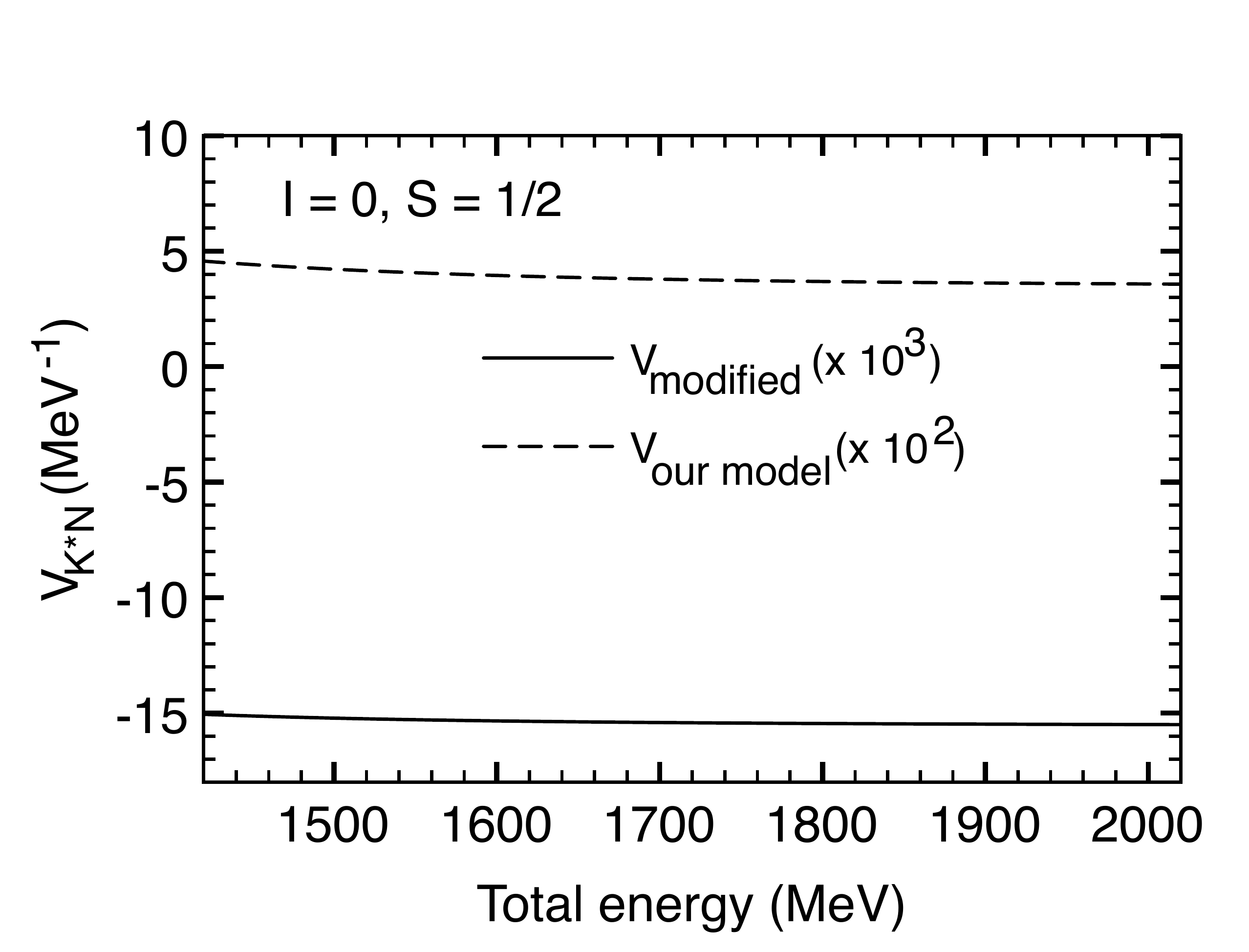}
\includegraphics[width=0.43\textwidth]{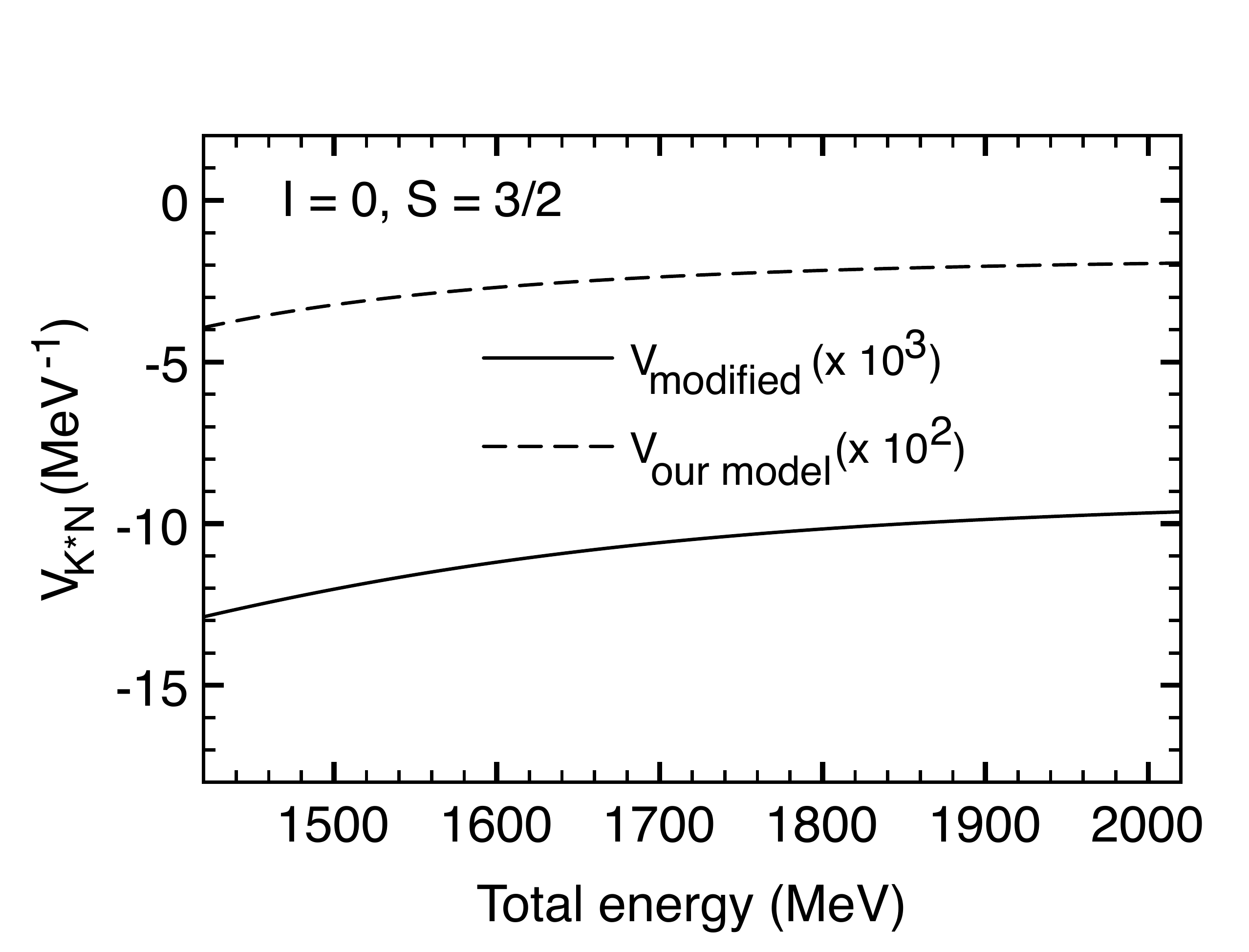}
\includegraphics[width=0.43\textwidth]{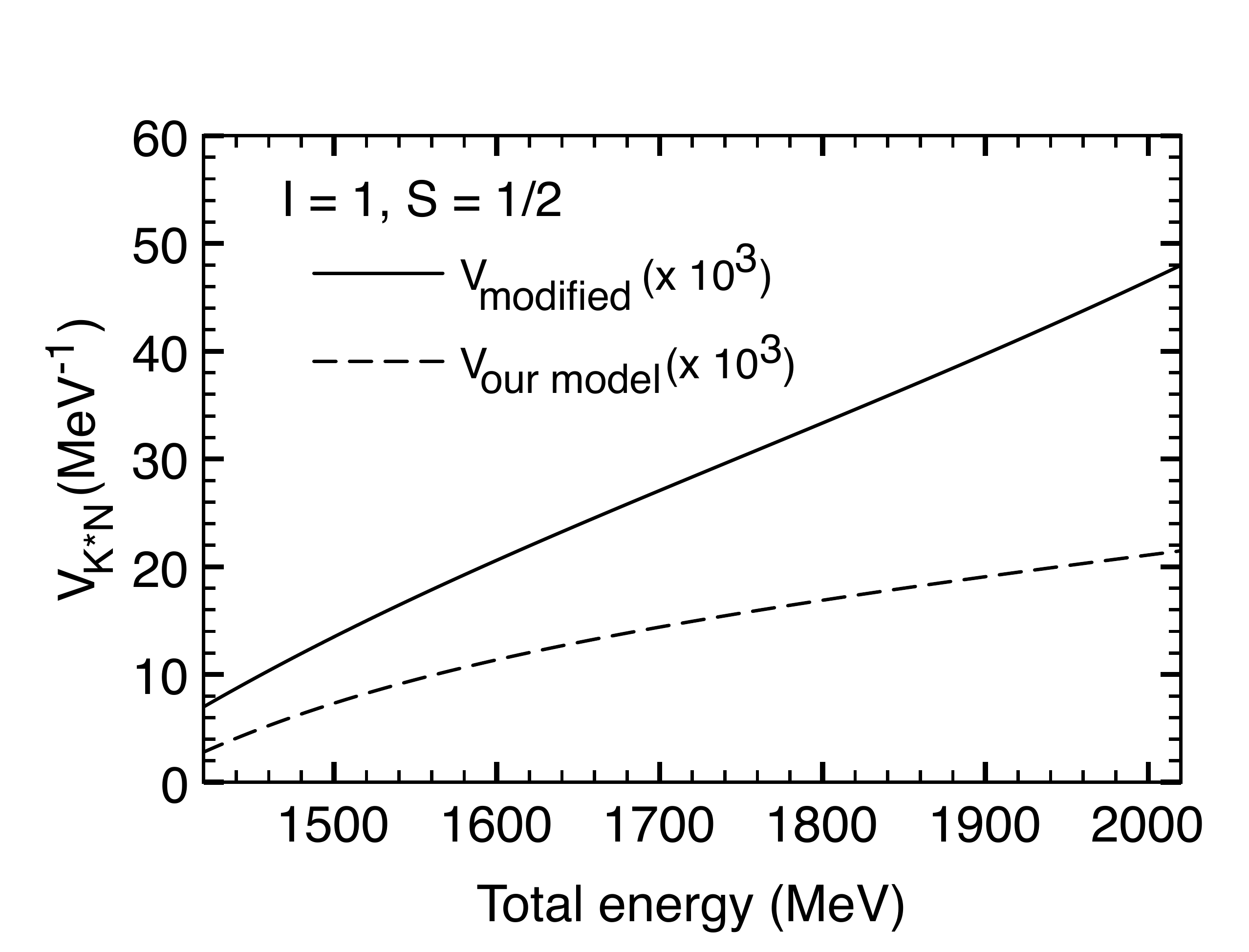}
\includegraphics[width=0.43\textwidth]{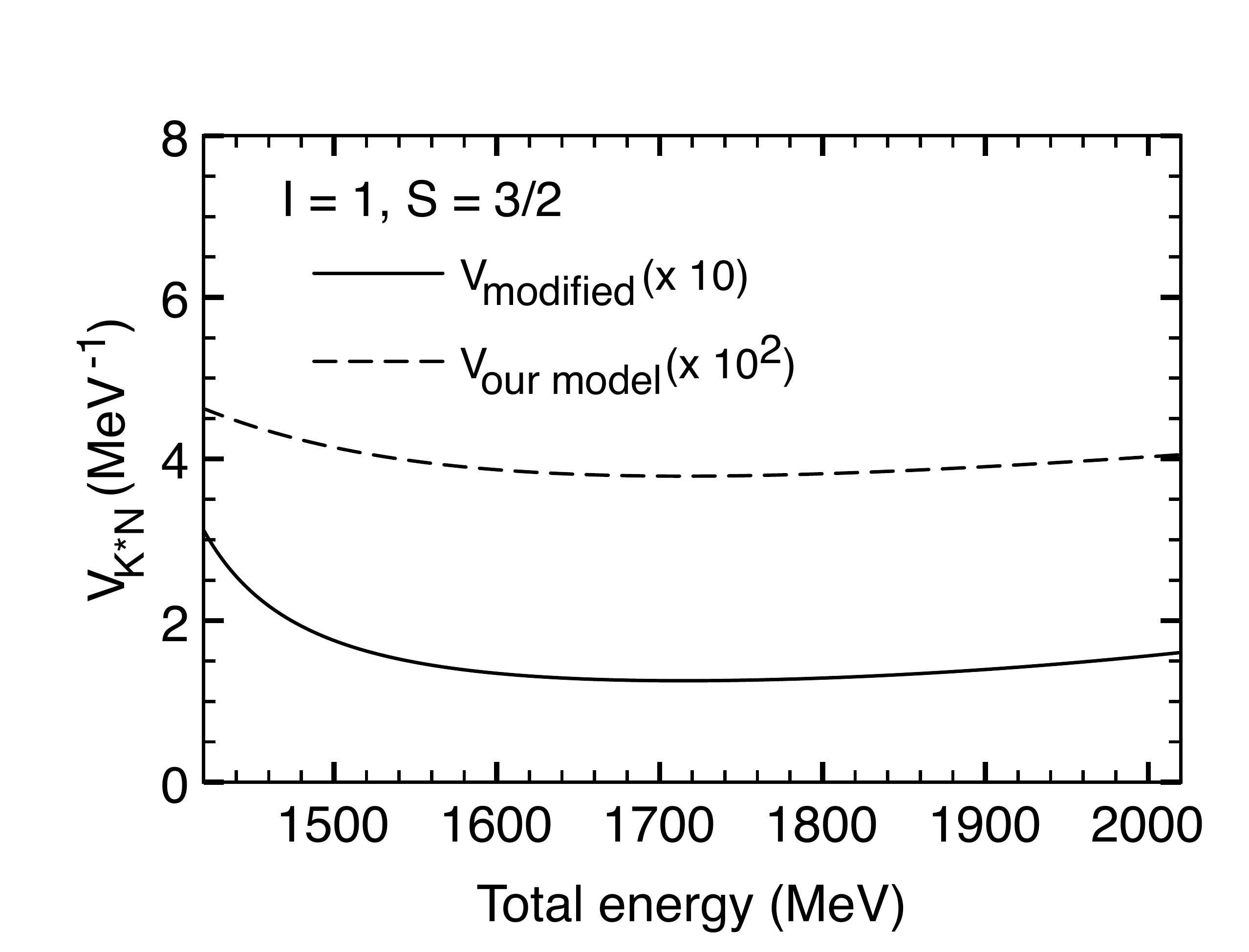}
\caption{Same as Fig.~\ref{fig:VKN} but for $K^*N$ system with $V_{\rm our~model}$ being the sum of the amplitudes obtained from Eqs.~(\ref{VKsN})  and (\ref{ampudia2}).  }\label{fig:VKstarN}
\end{figure}

For completeness, we compare the modified kernel and those obtained in Section~\ref{potentials} in Figs.~\ref{fig:VKN} and \ref{fig:VKstarN} for the $KN$ and $K^*N$ systems, respectively.

\begin{figure}[h!]
\begin{center}
\includegraphics[scale=0.28]{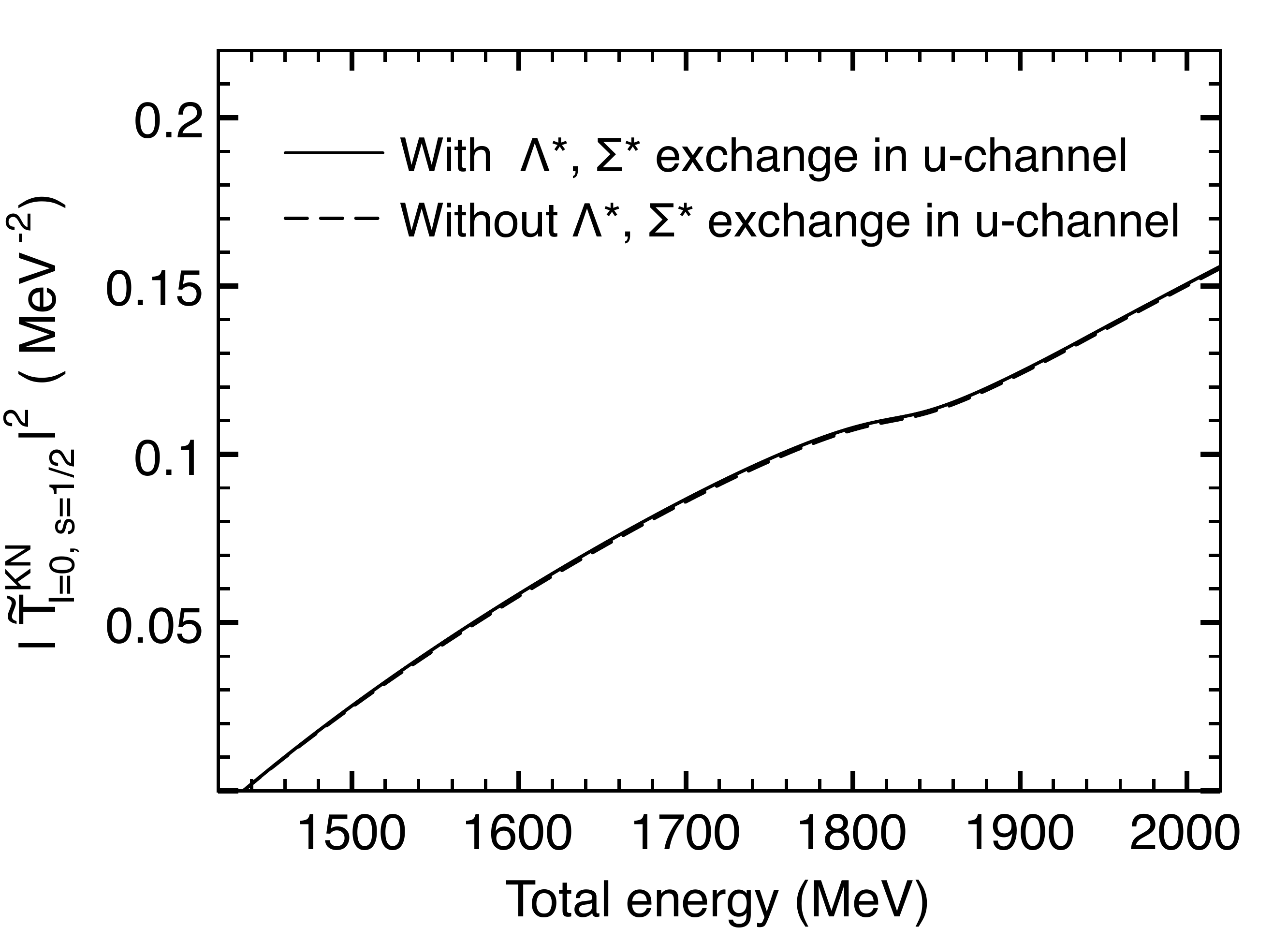}
\caption{Contribution of the resonance exchange in the $u$-channel diagram to the $KN$ amplitude in the $S_{01}$ partial wave obtained with parameter set II. Notice that the dashed and solid curves are almost indistinguishable. }\label{resinu}
\end{center}
\end{figure}

\begin{figure}[h!]
\centering
\includegraphics[width=0.45\textwidth]{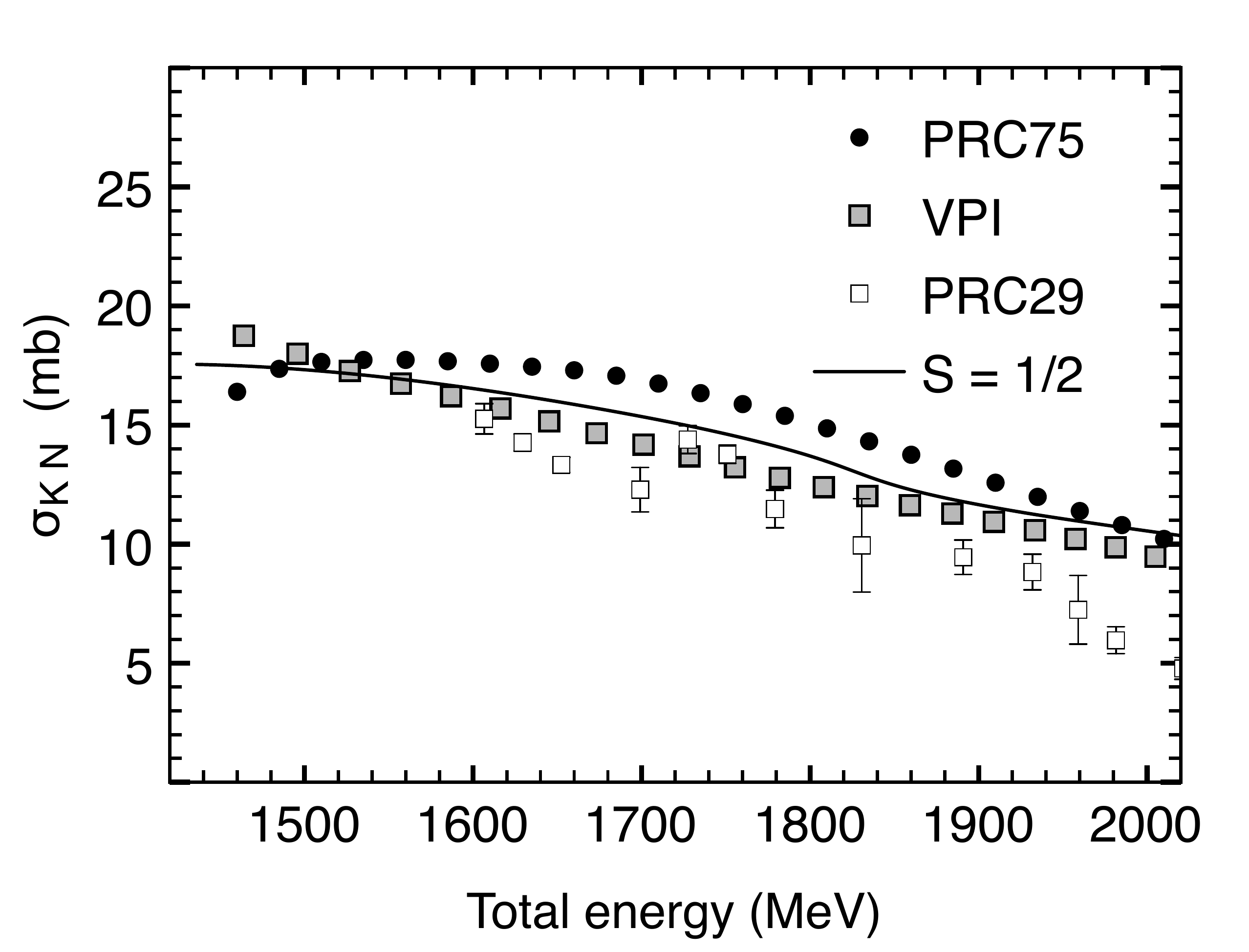}
\includegraphics[width=0.45\textwidth]{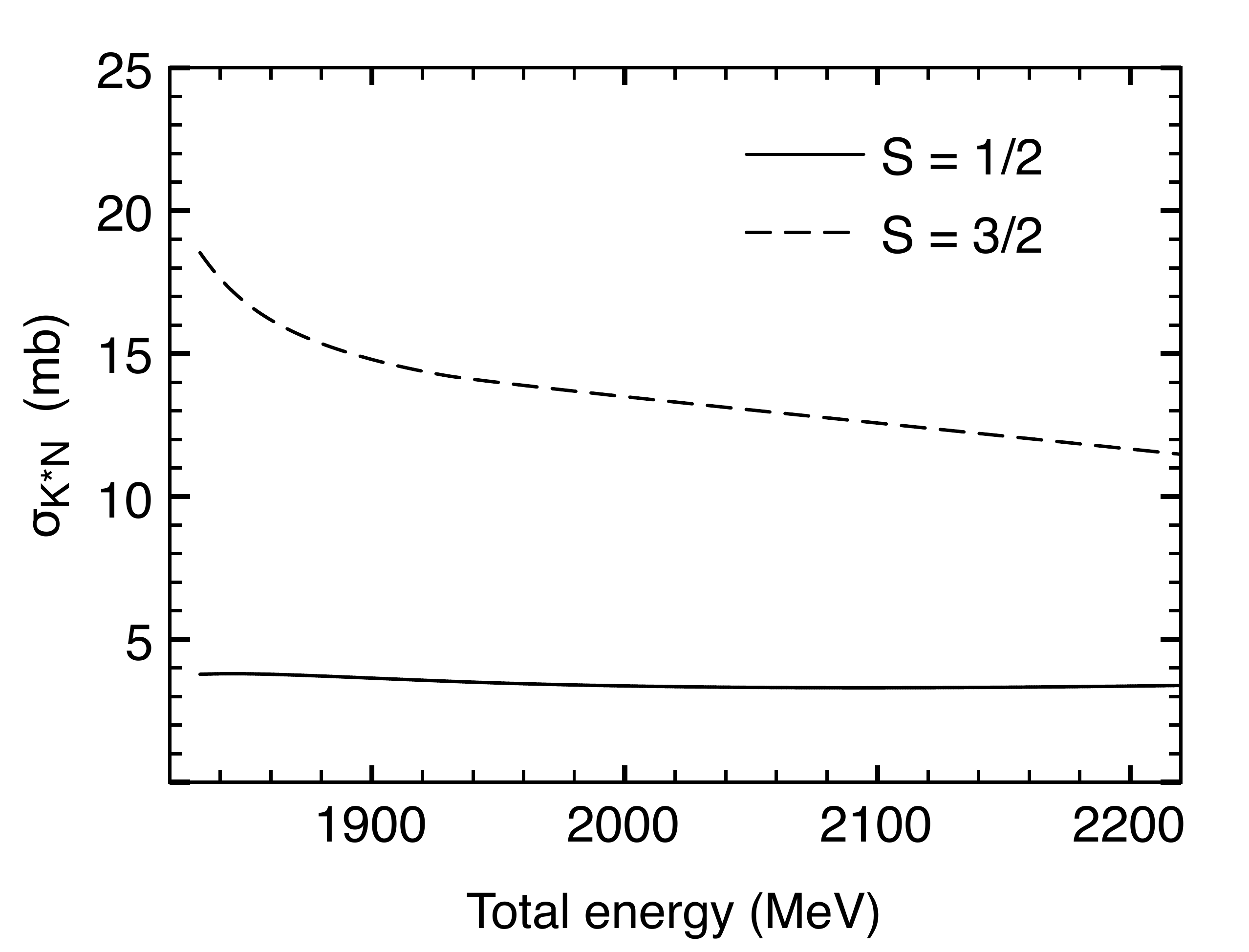}
\caption{Our results for the total cross sections  for the $KN$ (left panel) and $K^*N$ (right panel) $s$-wave interaction which leads to a total spin 1/2 (solid line) in the former case and  total spin 1/2 (solid line) and 3/2 (dashed line) in the latter one. The data represented by filled circles, boxes and empty triangles are taken from Refs.~\cite{prc75,gwu,prc29}, respectively.
}\label{three}
\end{figure}

The discussion made above and the comparison of the interaction kernels shown in Figs.~\ref{fig:VKN} and \ref{fig:VKstarN} indicates that some information is missing in our model.  A question that may arise at this point is if this missing information can be recovered by considering an exchange of a larger number of hyperon resonances in the $u$-channel diagrams. Although we already mentioned to have found 
the contribution obtained from resonance exchange to be negligibly small, to answer the aforementioned question, we find it useful to show this contribution qualitatively. For this we consider, as an example, the $S_{01}$ amplitude for the $KN$ system obtained with parameter set II. In Fig.~\ref{resinu} 
we show the results obtained for the isospin 0, spin 1/2 amplitude:  (1) without including any resonance exchange in $KN \to KN$, $K^*N \to K^*N$, $K N \leftrightarrow K^*N$  and, (2) by considering the resonances listed in Table~\ref{couplings} in all these processes. It can be seen in Fig.~\ref{resinu} that the contribution obtained from the resonance exchange in $u$-channel is insignificant. The case for isospin  1 is similar. These findings discourage us from considering  more  resonances, which implies increasing the number of parameters of the theory but obtain no significant contributions. It is possible that the missing contribution could be obtained by considering higher order terms in the Lagrangian. Although it would involve fixing a larger number of parameters using the data shown in Figs.~\ref{two} and \ref{three}, it would be an alternative approach and should be considered in the future.  Other possibilities can be considering box diagrams for all the processes, in line with the study of  VB systems in Ref.~\cite{javi,javi2}. The results presented in our manuscript could stimulate calculations of such corrections while serving as guidance for the findings of the experimental studies of  $K, K^*$ production in $p$-$p$ and $p$-A processes.

With a reasonable agreement  between our results and the available data on $KN$, we  search for poles in the complex plane.  We have calculated the amplitudes for isospin 0 and 1 and spin-parities $1/2^-$ and $3/2^-$. We do not find  any pole in none of these configurations which can be related to a physical state.  

It is interesting to add that a calculation of the $\pi KN$ system in $s$-wave led to the generation of a broad state with spin parity $1/2^+$ in Ref.~\cite{ourplb}. The present calculation is, in some sense, similar to the one made in Ref.~\cite{ourplb}, recalling that $K \pi N$ system can be reorganized as $K^*N$ too. However, the latter implies a $p$-wave interaction between the kaon and the pion.  This difference between the  kaon-pion interaction might be important for the formation of a resonance.

Next, we calculate the scattering lengths for the $KN$ system using the relation given by Eq.~(\ref{scatl}) and find $a^{I=0}_{KN} =-0.16$ fm and $a^{I=1}_{KN} =-0.29$ fm. The values found by different partial wave analyses groups for the $KN$ scattering lengths range from  $-0.105 \pm 0.01$~fm \cite{prc75} to $-0.23 \pm 0.18$~fm \cite{martin}, for isospin 0, and between $-0.286 \pm 0.06$~fm to $-0.308 \pm 0.003$~fm \cite{prc75}, for the isospin 1 case. 

Finally, we would like to present our results on the $K^*N$ system, which might be useful as an input to studies of $K^*$-mesons in hot and dense medium \cite{hades,star,NA49}. 
We show the total cross sections for spin 1/2 and 3/2 in the right panel of Fig.~\ref{three}, 
which have been calculated as
\begin{equation}
 \sigma_{S} = \frac{1}{2}\, \sigma_{I=0, S} + \frac{3}{2} \,\sigma_{I=1, S},
\end{equation}
 with
\begin{equation}
\sigma_{I, S} = \dfrac{1}{4\pi} \dfrac{M_N^2}{s} \mid T^{I,S}_{K^*N} \mid^2.
\end{equation}
The symbols $I$ and $S$ in the above equation represent the total isospin and spin, respectively, $M_N$ is the nucleon mass and $\sqrt{s}$ is the total   energy in the center of mass frame. This definition of an isospin averaged cross section is often used in literature \cite{prc75} and an analogous definition is used for isospin averaged amplitudes in the studies of mesons in medium \cite{Tolos:2010rm,Ilner:2013ksa,Cabrera:2014lca,Tolos:2008di}. 
Although we have already shown our results for the phase shifts for the $KN$ channel in Fig.~\ref{one}, for completeness, we show the corresponding cross sections too  in Fig.~\ref{three} (left panel). 

We also provide the scattering lengths calculated using Eq.~(\ref{scatl}) for the different spin-isospin configurations of the $K^*N$ channel in Table~\ref{slength}.

\begin{table}[h]
\caption{Scattering lengths for the $K^*N$ system.}\label{slength}
\centering 
\begin{tabular}{c|c|c|c|c}
\hline\hline
& $I=0$, $S=1/2$ & $I=0$, $S=3/2$&$I=1$, $S=1/2$&$I=1$, $S=3/2$\\\hline
$a^{I,S}_{K^*N}$ (fm) & (0.2,0.03) &(-0.08,0.04) &(0.1,0.0)&(-0.31,0.03) \\
\hline\hline
\end{tabular}
\end{table}

\section{Summary}
We can summarize the present work by mentioning that a coupled channel calculation involving both pseudoscalar and vector mesons has been done for strangeness $+1$ by taking 
different diagrams into account to obtain the kernel potential. In case of VB systems, we consider a contact term and the  $t$-, $u$-channel diagrams, with an exchange of an octet baryon or a light hyperon resonance for the latter one. For PB channels  we consider, in addition to the Weinberg-Tomozawa interaction, the $u$-channel exchange of light hyperon resonances. The exchange of hyperon resonances is found to give a negligible contribution. The subtraction constants required to regularize the loop have been fixed by  fitting the available data on the $KN$ phase shifts in the $S_{01}$ and $S_{11}$ partial waves. The resulting values of the subtraction constants turn out to be far from the natural values, indicating missing information on the interaction kernel in the model. We depict the kernels which are required to fit the data and discuss possible alternative calculations which can be done in future to obtain such interaction kernels.  With the calculations carried out in this formalism, we do not find any resonances. We present the total cross sections and scattering lengths for the $KN$ and $K^*N$ channels. Indeed, the results presented here are of special interest for $K$ and $K^*$ production in $p + p$ and $p + A$ collisions, as reported by HADES \cite{Agakishiev:2014nim,Agakishiev:2014moo,hades}, STAR \cite{star} and  NA49 \cite{NA49} Collaborations.

\section*{Acknowledgements} The authors thank Profs. D. Jido, A. Hosaka and H. Nagahiro for useful discussions.
The authors would also like to thank the Brazilian funding agencies FAPESP and CNPq for the financial support.  L.T  acknowledges support from the Ram\'on y Cajal
Research Programme (Ministerio de Ciencia e Innovaci\'on) and from  Grants No. FPA2010-16963 and No.  FPA2013-43425-P (Ministerio de Ciencia e Innovaci\'on), No. FP7-PEOPLE-2011-CIG under Contract No. PCIG09-GA-2011-291679 and the European Community-Research Infrastructure Integrating Activity Study of Strongly Interacting Matter (acronym HadronPhysics3, Grant Agreement n. 283286) under the Seventh Framework Programme of EU.

\end{document}